\documentclass[twocolumn, amsmath,amssymb, rmp]{revtex4}
\usepackage{graphicx}
\usepackage{dcolumn}
\usepackage{bm}
\usepackage{natbib}

\newcommand{\be}{\begin{equation}}
\newcommand{\ee}{\end{equation}}

\begin{document}

\title{Shear thickening in concentrated suspensions: phenomenology, mechanisms, and relations to jamming}
\author{Eric Brown}
\affiliation{School of Natural Sciences, University of California, Merced, CA 95343}
\author{Heinrich M. Jaeger}
\affiliation{James Franck Institute, The University of Chicago, Chicago, IL 60637}

\date{\today}

\begin{abstract}

Shear thickening is a type of non-Newtonian behavior in which the stress required to shear a fluid increases faster than linearly with shear rate.  Many concentrated suspensions of particles exhibit an especially dramatic version, known as Discontinuous Shear Thickening (DST), in which the stress suddenly jumps  with increasing shear rate and produces solid-like behavior.  The best known example of such counter-intuitive response to applied stresses occurs in mixtures of cornstarch in water. Over the last several years, this shear-induced solid-like behavior together with a variety of other unusual fluid phenomena has generated considerable interest   in the physics of densely packed suspensions.  In this review, we discuss the common physical properties of systems exhibiting shear thickening,  and different mechanisms and models proposed to describe it.  We then suggest how these mechanisms may be related and generalized, and propose a general phase diagram for shear thickening systems.
We also discuss how recent work has related the physics of shear thickening to that of granular materials and jammed systems.  Since DST is described by models that require only simple generic interactions between particles, we outline the broader context of other concentrated many-particle systems such as foams and emulsions, and explain why DST is restricted to the parameter regime of hard-particle suspensions.  Finally, we  discuss some of the outstanding problems and emerging opportunities.

 \end{abstract}
 
 
 \maketitle
 
 \section{Overview}

In a fluid, the energy dissipation rate under shear is characterized by the viscosity, defined as the ratio of shear stress to shear rate during steady flow. For Newtonian liquids in which the molecules interact in thermodynamic equilibrium, the viscosity is an intrinsic material parameter and is independent of the shear rate.   Suspending small particles in a Newtonian liquid can bring about  non-Newtonian behavior.  In this case viscosity may vary with shear rate, and in the field of rheology, where the term viscosity is used more broadly,  the viscosity is usually given as a function of shear rate.   However, the extension of the term viscosity to non-Newtonian fluids comes at the cost of some of the generality of the Newtonian viscosity.   Non-Newtonian fluids with large or densely-packed  particles may be non-ergodic, meaning the particle arrangements are not necessarily in thermal equilibrium, but instead can exhibit hysteresis.  Thus, while the viscosity ideally defines a local relationship between shear stress and shear rate that is valid everywhere in a fluid, this is not always the case for non-Newtonian fluids. 


In some instances the energy dissipation rate decreases with increasing shear rate, resulting in behavior labeled shear thinning. For applications such as paints, this is desirable as it lets pigments flow easily when brushed but minimizes drips when there is no brushing action. The opposite type of non-Newtonian flow behavior,  in which the energy dissipation rate increases with shear rate, is shear thickening.  

 \begin{figure}
\includegraphics[width=3.4in]{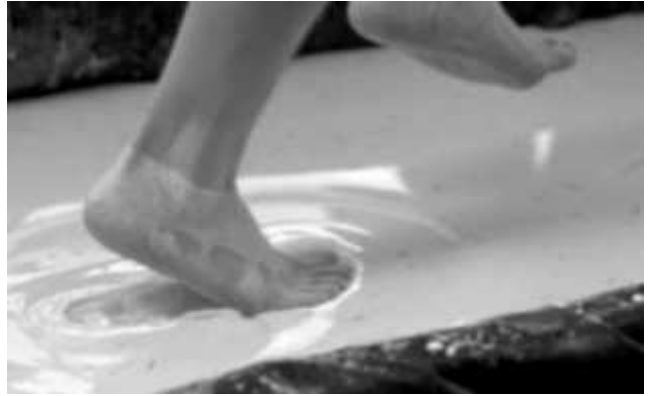}
\caption
{
Snapshot of a person running on top of a pool filled with a dense suspension of cornstarch and water.  The fluid can temporarily hold up the person's weight like a solid, sustaining stresses orders of magnitude beyond the capabilities of the suspending Newtonian liquid (here water).  The suspension behaves liquid-like before and after impact, for example gravity waves are seen to propagate along the surface, and without running, the person would sink into the suspension as in a normal liquid.  
}
\label{fig:cornstarchpool}
\end{figure}

To most observers, the whole notion of a liquid becoming thicker when stirred or sheared is utterly counter-intuitive.  Perhaps the best known, and certainly the most widely available, material to exhibit dramatic thickening is a densely packed suspension of cornstarch particles in water.  Such suspensions are sometimes referred to as discontinuous shear thickeners because of the apparently discontinuous jump in the viscosity with increasing shear rate.   Such a suspension feels like a liquid at rest, but  when stressed or sheared, its resistance to flow increases dramatically and it appears to take on solid-like properties; for example a person can run across the surface of a pool filled with a suspensions of cornstarch in water without sinking in (Fig.~\ref{fig:cornstarchpool}).  Other notable solid-like phenomena observed in dense, shear thickening suspensions include cracking of the fluid under impact \citep{RMJKS13}, and the formation of stable macroscopic structures under vibration \citep{MDGRS04}.  This shear thickening is completely reversible; once the stress is removed, the suspensions relax and flow like any other liquid.  

The remarkable increase in flow resistance of shear thickening suspensions can cause problems in their industrial processing, such as jamming when they are extruded through small openings, or even breaking of mixing equipment \citep{Ba89}.  On the other hand, one useful property of such suspensions is that they can provide remarkably effective energy dissipation.  This has opened up new opportunities for use in flexible protective gear ranging from sports padding to stab-proof vests that protect the wearer by becoming semi-rigid in response to impact, while otherwise remaining fluid-like and flexible to allow freedom of movement \citep{LWW03, JGXJYLL13, PH13}.   

Despite the fact that shear thickening is typically referred to as less common than its counterpart, shear thinning, it has  in fact been observed across very a broad range of  colloidal and non-colloidal suspensions, with hundreds of publications  in the scientific literature  since the 1930's.  It was even suggested by  H. A. Barnes in his influential 1989 review that perhaps all suspensions could exhibit shear thickening under the right conditions\citep{Ba89}.  Consequently, one of the key problems in the field since then has been to develop general models.   

What makes the thickening phenomenon particularly intriguing from a fundamental science perspective is that all its hallmarks can be exhibited already by the very simplest type of suspension, namely hard spheres suspended in a Newtonian liquid.  However, it has been a challenge to understand why strong shear thickening tends to be observed in densely packed suspensions of non-attractive hard particles, but has not been observed in suspensions of attractive particles, or complex fluids consisting of soft particles. This has been particularly puzzling, since proposed mechanisms have focused on very general features such as microstructural changes under shear with corresponding changes in lubrication forces and other interactions between particles \citep{Ho74, BB85, BJ12}.  Understanding how such a simple combination of ingredients  leads to such dramatic behavior and developing quantitative predictions that can apply to such a wide range of suspensions, while accurately identifying the parameter regime under which shear thickening is observed has remained a fundamental problem that modern models of shear thickening have tried to address. 

 

In the time before Barnes' review, shear thickening was studied mainly in the chemical engineering community, motivated  by multiphase fluid processing problems.  Since then, there has been renewed interest from the soft condensed matter physics community, much of it motivated by connecting shear thickening systems to other soft matter systems through the universal concept of jamming.   In jamming, a fluid-solid phase transition occurs due to system-spanning networks of particle contacts when the particle density increases \citep{CWBC98, LN98}.  This and other concepts coming from granular physics have injected new ideas into the field and have led to  revised models for shear thickening.  

Furthermore, over the last decade new experimental techniques have been developed that can probe the dynamical structure of the particle subphase in-situ, while the suspension is being sheared.  This includes x-ray and neutron scattering techniques \citep{MW02} as well as direct observation of individual particles by confocal microscopy \citep{CMIC11}.  With increases in computational power, simulations can now deal with significant particle numbers and system sizes \citep{MB04a, WB09, NM12}.   These advances have made it possible to investigate microstructural changes at the onset of shear thinning or thickening and thus test models at the particle level.   
This review focuses mostly on the progress since Barnes' review, emphasizing discontinuous shear thickening and dynamic phenomenology in dense suspensions.   While much progress has been made, there is still  contention in the literature about the mechanism(s) that are responsible for the observed dramatic increases in shear stress in discontinuous shear thickening.  Recent publications are roughly evenly split in attributing this shear thickening to three different mechanisms.  One mechanism is hydroclustering, where particles tend to push together into clusters under shear and this rearrangement leads to increased lubrication drag forces between particles \citep{BB85, WB09}.  A second mechanism is an order-disorder transition, in which the flow structure changes from ordered layers to a disordered structure, which also results in an increase in drag forces between particles \citep{Ho74}.  A third mechanism is dilatancy, in which the volume of the particulate packing increases under shear, which pushes against the boundaries and  can result in additional stresses from solid-solid friction \citep{BJ12}.  With this review, we hope to bring the community closer to a consensus on the role and range of applicability of each of these mechanisms  \footnote{full disclosure: the authors have been proponents and developers of the dilatancy mechanism in their previous work}.  We will discuss these mechanisms and the issues of contention in detail in later sections, but some of the issues are worth bringing up now.

One of the major issues  is whether the hydroclustering mechanism, which accurately describes weak shear thickening in less densely packed suspensions, can also explain discontinuous shear thickening and associated phenomena in more densely packed systems.  It seems likely that hydroclustering is a trigger for the onset of shear thickening,  yet the incredibly high stresses observed in discontinuous shear thickening are too large to be explained by lubrication forces. They appear to be  better explained by a fabric of stress paths that span the system and support normal stresses of similar magnitude as the shear stress, a situation familiar from granular matter. An interesting consequence for dense, hard sphere suspensions  is that boundaries play a critical role and extremely strong thickening is not an intrinsic bulk material response.  Hand in hand with this behavior likely come other `granular' features, such as an inherent heterogeneity and propensity for strain localization (`shear banding').  To sort this out, more attention will need to be focused on the connections between local and global rheology in these systems.

A second issue concerns particle size.  Can the same mechanisms explain shear thickening in both Brownian colloids and non-Brownian suspensions despite the fact that the particle microstructure dynamics are very different?  Traditionally,  the prevailing answer has been no.  We have recently introduced a new perspective, focused on consideration of the dominant stress scales,  which suggests that in many situations  Brownian and non-Brownian suspensions can be understood in similar terms \citep{BFOZMBDJ10,BJ12}.


This review will also cover some of the dynamic, non-steady-shear phenomena commonly associated with shear thickening and discuss to what extent they are related to standard, steady-state shear thickening.  One example are the stable macroscopic structures that emerge under vibration, which can be attributed to hysteresis in the viscosity curve \citep{De10}.  Another example concerns the  explanation of the most dramatic behavior commonly associated with shear thickening -- why a grown person does not sink in when jumping onto or running across a dense suspension of, e.g., cornstarch in water \citep{WJ12}.  Recent results revealed that this phenomenon is closely connected to a dynamic form of jamming but not the same response as steady state shear thickening.   The fact that these realizations have been made only in the past few years highlights that, historically, the vast majority of work on non-Newtonian thickening has focused on simple, steady state shear.  The recent influx of interest in jamming from soft matter physics has been a main source of motivation for investigation of dynamic or transient jamming phenomena, and this trend may lead to new insights also into well-known suspension phenomena. At the same time it opens up new opportunities to explore  jamming in a different context.

This review is structured as follows.  In Sec.~\ref{sec:rheology}, we introduce some basic concepts and definitions of rheology.  Sec.~\ref{sec:characterization} defines and characterizes the various types of shear thickening that have been reported in the literature.   In Sec.~\ref{sec:mechanisms} we outline the main mechanisms that have been proposed  to explain shear thickening.     Sec.~\ref{sec:stressscales}  shows some of the important scalings of stresses in shear thickening, and discusses how much of the  rheology can be interpreted in terms of dominant stress scales.  A  state diagram is introduced that delineates the region of observable shear thickening behavior.   In Sec.~\ref{sec:jamming}, we broaden the scope and discuss connections to jamming as well as granular materials and other many-particle complex fluids, including foams and emulsions.  Sec.~\ref{sec:microstructure}  discusses in detail the consequences of the fact that the rheology can be determined by boundary conditions and a global structure, as opposed to an intrinsic rheology determined by the local microstructure.  Dynamic phenomena  often associated with shear thickening are described in Sec.~\ref{sec:phenomenology}.  We close with a summary and some key outstanding issues and opportunities.

 \section{Introduction to rheology}
 \label{sec:rheology}
 
 \begin{figure*}
\includegraphics[width=5.in]{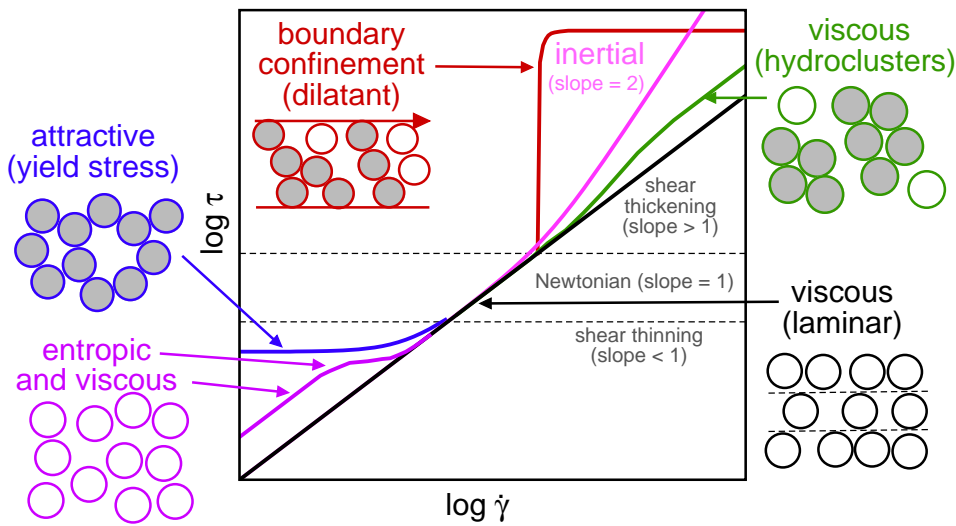}
\caption
{
Sketch of different possible regimes of shear stress $\tau$ vs.~shear rate $\dot\gamma$ for suspensions, plotted on a log-log scale.  Various contributions to stresses and their associated particle arrangements are indicated for different regimes of flow response to applied shear:  shear thinning, Newtonian, and shear thickening.  A particular complex fluid may exhibit several of these regimes, depending on  material properties and dominant forces.  Figure based on \citet{BJ11}.
}
\label{fig:stress_structures}
\end{figure*}


The viscosity $\eta$ of a complex fluid relates the  shear stress $\tau$ in a steady flow to the shear rate $\dot\gamma$ via $\tau = \eta \dot\gamma$.  Some examples of different types of $\tau(\dot\gamma)$ are sketched in Fig.~\ref{fig:stress_structures}.   On  a log-log scale, a Newtonian regime has slope 1, shear thinning regimes corresponds to slopes $\partial\log\tau/\partial\log\dot\gamma <1$, and shear thickening regimes correspond to slopes $\partial\log\tau/\partial\log\dot\gamma >1$.  Depending on the types of particles and suspending fluids and their material parameters, different regimes are observed in different ranges of shear rate.  We note that a single complex fluid may exhibit several regimes and can exhibit both shear thinning and shear thickening.  On the other hand, depending on the material parameters, not all suspensions exhibit all regimes.  

An approximation that is useful for interpretation of rheological curves is that interparticle forces originating with different mechanisms  add together to produce the net relation between shear stress and shear rate.  For example, many complex fluids exhibit a yield stress, corresponding to a critical stress that must be applied before the shear rate becomes non-zero.  This results from stable static structures which can be due to a number of different forces including interparticle attractions \citep{TPCSW01}, repulsions from an electrostatic potential \citep{MW01a}, steric (solid particle) repulsion \citep{Ho98, OSLN03}, gravitational pressure \citep{FBOB09, BJ12}, and attractions from induced electric or magnetic dipoles \citep{BFOZMBDJ10}.  The scale of the yield stress can be estimated as the scale of the force between neighboring particles divided by the cross-sectional area of a particle when the particle packing density is high enough to support percolating structures. These fluids exhibit strong shear thinning beyond the yield stress as the total shear stress remains largely independent of shear rate, and may exhibit Newtonian or shear thickening behavior at higher shear rates once other sources of stress exceed the yield stress.  

If the particles in a complex fluid are smaller than about 1 micron,  they experience Brownian motion and an effective repulsive pressure from effective entropic forces.  Such fluids may have a Newtonian regime at low shear rates, followed by shear thinning, followed by a second Newtonian regime at higher shear rates.  Shear thickening may result at higher shear rates if some mechanism produces stresses that increase faster than linearly with shear rate.  The different types of shear thickening will be discussed in more detail in Sec.~\ref{sec:characterization}.  Despite the differences among different complex fluids, in all cases the broader problem in rheology consists of how to attribute fluid properties to particle interactions and material properties on the one hand and microstructural changes in the fluid on the other.  

In the above examples of Newtonian and shear thinning behavior, as well as hydrocluster-based shear thickening, the relationship between shear stress and shear rate is understood to be true locally at every point in the fluid.   However, the situation turns out to be more complicated for other types of apparent shear thickening: inertial and dilatant.  The relationship between shear stress and shear rate is traditionally obtained by a rheometer which measures the drag force on a moving surface at the boundary of the fluid as a function of the tool velocity.  The local relationship between shear stress and shear rate can be inferred based on the geometry of the flow region, using the assumptions that the fluid is homogeneous, the flow profile is laminar, and the local shear stress is purely a function of the local shear rate.  It turns out that these assumptions are  violated for inertial and dilatant shear thickening, with significant consequences for the interpretation of the behavior.  This issue will be revisited in Sec.~\ref{sec:intrinsic}, following a more detailed descriptions of the different types of shear thickening and the mechanisms proposed to describe them.

 \section{Characterization of different types of shear thickening}
\label{sec:characterization}

Shear thickening is technically a category of non-Newtonian behavior, corresponding to any rheology in which the effective viscosity increases with shear rate.  Since there are several different types of shear thickening, each characterized by certain defining features and likely due to different mechanisms,  we first summarize the basic features of some of the main types of shear thickening reported.

\subsection{Continuous shear thickening}

The degree to which the viscosity increases with shear rate depends on the volume fraction of solid particles, also referred to as the packing fraction $\phi$. Shear thickening is generally not observed in dilute suspensions, but starts to gradually appear at intermediate packing fractions, typically around $0.3 \stackrel{<}{_\sim} \phi \stackrel{<}{_\sim} 0.4$ for suspensions of solid spheres \citep{MB04a, WB09, NM12}. At these particle concentrations, the viscosity increase is relatively mild, perhaps up to several tens of percent over the few decades of shear rates observed in typical experimental settings (see the green curve in Fig.~\ref{fig:stress_structures} for a comparison to other types of shear thickening). This type of shear thickening is often referred to as `continuous'.  The rate of increase in viscosity with shear rate gradually becomes larger with increasing packing fraction, and it is usually found that the shear thickening regime starts at a critical stress $\tau_{min}$ which is roughly independent of packing fraction \citep{Laun84, MW01a, GZ04, WB09}.  Below this stress, shear thinning or a Newtonian regime may be found, depending on the suspension.


\subsection{Discontinuous shear thickening}
\label{sec:DSTdef}

\begin{figure}                                                
\centerline{\includegraphics[width=3.4in]{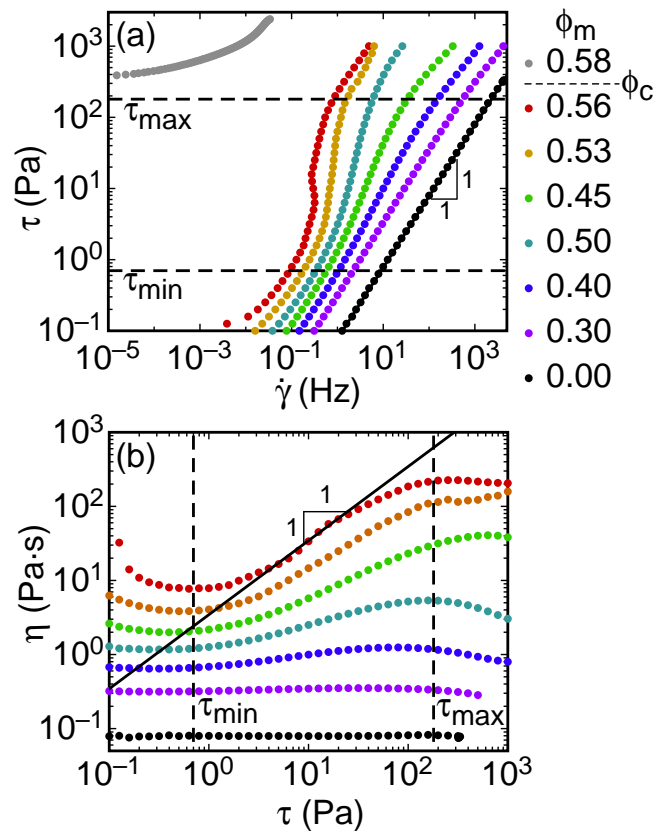}}
\caption{Representative viscosity curves showing the evolution of DST with increasing packing fraction.  The suspension consists of cornstarch in a solution of 85\% glycerol and 15\% water by weight, with different mass fractions $\phi_m$ (proportional to $\phi$) shown in the key.  (a) Shear stress $\tau$ vs. shear rate $\dot\gamma$, in which shear thickening is defined by the region with slope greater than 1.  The evolution to apparently discontinuous viscosity curves  can be seen as the packing fraction increases toward $\phi_c$.  Above $\phi_c$, the suspension becomes a yield stress fluid. (b) Same data, replotted as viscosity $\eta$ vs $\tau$.  The latter format better shows the gradual evolution of the increasing slope in the shear thickening regime, confined in the stress range between $\tau_{min}$ and $\tau_{max}$.  When plotted as $\eta(\tau)$, a slope greater than zero corresponds to shear thickening, and a slope of 1 corresponds to a discontinuous jump in $\tau(\dot\gamma)$.  Figure based on \citet{BJ12}.}
\label{fig:visc_curves}                             
\end{figure}

In many shear thickening fluids, the viscosity increase with shear rate continues to become steeper with increasing packing fraction, up to the point that the viscosity and shear stress appear to jump discontinuously by orders of magnitude beyond a certain shear rate (such as the red curve in Fig.~\ref{fig:stress_structures}).  In such cases it is often said that the shear thickening evolves from continuous to discontinuous shear thickening with increasing packing fraction.  So-called Discontinuous Shear Thickening (DST) has the most dramatic increase in viscosity of any type of shear thickening, and includes the prototypical example of cornstarch in water as well as many other densely packed hard-particle suspensions \citep{MW58, Ho72, Ho74, Ho82, Ba89, BLS90, Laun94, FHBM96, BW96, OM00, MW01a, MW01b, MW02, BBS02, LDH03, LDHH05, EW05, SW05, ENW06, LW06, FHBOB08, BJ09, BFOZMBDJ10, BZFMBDJ10}, and solutions of micelles \citep{HRH91, LP96}.  An example of the evolution from continuous to discontinuous shear thickening with packing fraction is shown in Fig.~\ref{fig:visc_curves}. 

The DST suspensions mentioned above tend to have several rheological properties in common that  provide considerable insight into the possible mechanisms and help distinguish different phenomena.  One such property is that the DST regime tends to occur in a well-defined range of shear stress.  The onset of the shear thickening regime can be characterized by the same critical stress $\tau_{min}$ that is roughly independent of packing fraction as with continuous shear thickening (see Fig.~\ref{fig:visc_curves}b).  Once started, the viscosity or shear stress increase does not continue indefinitely with increasing shear rates. Instead, the shear thickening regime ends at a maximum stress scale $\tau_{max}$, also roughly independent of packing fraction \citep{FHBM96, MW01a, SW05, BJ09}.  Above this stress, shear thinning, cracking, and breakup of the suspension are often observed \citep{Laun94}.

A second common property has to do with the scaling of the slope of $\tau(\dot\gamma)$ in the shear thickening regime.  The apparently discontinuous jump in the viscosity or shear stress with shear rate tends to be observed only over a range of packing fractions of a few percent in very densely packed suspensions, typically around $\phi \approx 0.6$ for nearly spherical particles \citep{MW01a, EW05, BJ09}.   This critical packing fraction corresponds to the jamming transition, above which the system has a yield stress like a solid \citep{LN98}.  The value of $\phi_c$ can vary with particle shape and a number of other suspension properties, but the proximity to this point generally controls the slope of shear thickening regime like a second order phase transition; the slope of $\tau(\dot\gamma)$ diverges at $\phi_c$ \citep{BJ09, BZFMBDJ10}.  This critical point will be discussed in more detail in Sec.~\ref{sec:jamming}.   



\subsubsection{Local vs global descriptions of rheology and nomenclature}
\label{sec:intrinsic}

There is a major distinction between the local relation between shear stress and shear rate and the energy dissipation rate measured by a rheometer for DST suspensions just described.  It has been found that the local shear stresses between neighboring particles are frictional and thus proportional to the local normal stress, which can depend on the boundary conditions, and the global structure of a transiently jammed system is required to sustain these contacts (to be discussed in detail in Sec.~\ref{sec:normalstress}).  These contributions can be separated from purely hydrodynamic contributions to the shear stress using modern optical microscopy and high-precision normal-force-controlled measurements \citep{BJ12}.  In this situation, the local shear stress is not simply a function of the shear rate.  When these global effects are separated from the local  relationship between shear stress and shear rate, surprisingly the local $\tau(\dot\gamma)$ relationship can be Newtonian or even shear thinning.  This is possible, because all of the stresses that were responsible for shear thickening originate from non-local structure and boundary conditions, and the direct local dependence is on the normal stress which is not linked to shear rate \citep{BJ12}.  This was found to be true for the prototypical shear thickener cornstarch in water, among other DST suspensions.  


As a result, a subtle problem emerges with the nomenclature for DST.   While the modern trend in the rheology community has been to define rheology based solely on  local stress relationships, only recent technological improvements in measuring techniques have allowed experiments to observe local structure and directly test  local relationships. The literature going back to the 1930's refers to shear thickening based on global rheometer measurements from which the local stresses and shear rates were inferred based on assumptions that are now known to be over-simplified. Therefore, the term `shear thickening' is technically incorrect when applied to  the majority of the existing literature on the subject, including the prototypical cornstarch in water suspension.  

To avoid  a major revision of the existing literature, we prefer to keep the term `discontinuous shear thickening', but use it in capitalized form `Discontinuous Shear Thickening' (DST). This allows us to put a label to a phenomenon that has the well-defined characteristics outlined in Sec.~\ref{sec:DSTdef}, but is not necessarily `shear thickening' in terms of a local $\tau(\dot\gamma)$ relationship.  Strictly speaking, even the term `discontinuous'  may not be an accurate description of the rheological curve, since when $\tau$ is the control parameter it can be seen that the viscosity curve evolves continuously (Fig.~\ref{fig:visc_curves}) and only becomes discontinuous in the limit of $\phi_c$ \citep{BJ09, BZFMBDJ10}.  

The use of a proper name rather than the generic descriptor `discontinuous shear thickening' also helps distinguish this phenomenon from other types of shear thickening that may appear discontinuous, but have  properties that would suggest different mechanisms. Examples include dramatic irreversible shear thickening, in some cases attributable to chemical-attraction-induced aggregation \citep{OKW08, LKZW10}.  Others occur only as transient behavior \citep{FLBBO10} or require the application of strong electric or magnetic fields \citep{Ti11}.

 \subsection{Inertial effects}
 \label{sec:inertia}
 
One type of apparent shear thickening is characterized by a scaling $\tau(\dot\gamma) \propto \dot\gamma^2$ in the limit of high shear rates. First described by Bagnold \citep{Ba54}, it has since been reported in a wide variety of rheology experiments including different flow geometries \citep{HZCB02,FLBBO10}.  The dependence on packing fraction is relatively weak compared to discontinuous shear thickening, and tellingly, this behavior can even be observed in pure Newtonian liquids (zero particle packing fraction) at high shear rates \citep{BJ12}.  While various detailed descriptions have been given for this scaling behavior, in general it can be attributed to inertial effects such that the force required to displace a mass of material scales as velocity squared in the limit of high speeds.    The transition from a Newtonian scaling ($\tau\propto \dot\gamma$) to an inertial scaling ($\tau(\dot\gamma) \propto \dot\gamma^2$) has been characterized in terms of a Reynolds number, Bagnold number, or Stokes number, which are all equivalent in terms of their scalings with $\dot\gamma$, although the prefactors vary.  There can also be a scaling regime  where $\tau\propto \dot\gamma^{3/2}$ that exists for a partially inertial flow at intermediate Reynolds numbers between about $10^0$ and $10^3$ \citep{BJ12}.  


Perhaps the most developed and general description for this inertial behavior is in terms of the Reynolds number and turbulence.  Here the inertial scaling ($\tau\propto \dot\gamma^2$) comes from  the momentum advection term in the Navier-Stokes equations, which is dominant  for any fluid at high shear rates and Reynolds numbers.  In this regime, inertia leads to flow instabilities which result in a non-laminar flow and can include counter-rotating eddies on various scales.  These eddies increase the local shear rate within the fluid compared to a laminar flow and are responsible for the increased rate of energy dissipation, which explains why this behavior can be observed even for a pure Newtonian liquid.    Since it is widely known that the increase in energy dissipation rate occurs because the flow profile becomes non-laminar, rather than because of a change in the  local $\tau(\dot\gamma)$ relationship, these inertial effects are typically not referred to as shear thickening. However, this convention is not universal and in some publications such effects are still referred to as shear thickening \citep{Ba54, FLBBO10}.  



\subsection{Distinction between different types of shear thickening}
   
 One practical way to distinguish different types of shear thickening is by fitting a power law $\tau \propto \dot\gamma^{\alpha}$ to obtain the exponent $\alpha$.  A Newtonian flow corresponds to $\alpha=1$.  Inertial effects correspond to $\alpha=2$ in the limit of high shear rates, independent of packing fraction.  DST is characterized by large $\alpha$ which approaches infinity as the critical packing fraction $\phi_c$ is reached.  On the other hand, continuous shear thickening is typically characterized by $\alpha$ only slightly larger than 1, approaching 1 in the limit of zero packing fraction.  While there is not a sharp transition between continuous and discontinuous shear thickening ($\alpha$ increases continuously with packing fraction), the former usually evolves into the latter as the packing fraction is increased. In practice, systems are often referred as to discontinuous if $\alpha \stackrel{>}{_\sim} 2$ and increasing with packing fraction, and continuous if $1< \alpha  \stackrel{<}{_\sim}  2$.   
 


\section{Proposed mechanisms}
 \label{sec:mechanisms}
 
\subsection{Hydroclustering}

\begin{figure}                                                
\centerline{\includegraphics[width=3.4in]{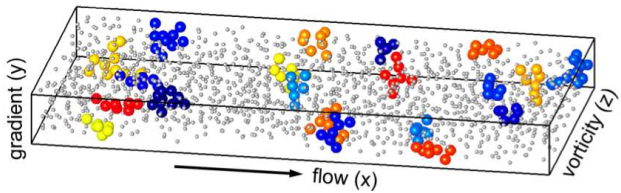}}
\caption{Instantaneous configurations of transient clusters  in the shear thickening regime, observed using fast confocal rheology.  Different colors indicate different clusters.  Particles outside the large clusters are drawn with smaller size for clarity.  Figure reproduced from \citet{CMIC11}. }
\label{fig:hydroclusters}                             
\end{figure}

The hydrocluster mechanism was first introduced by Brady and Bossis in 1985 \citep{BB85}.  The basic concept is that particles are pushed into each other by shear, and to move away from each other they must overcome the viscous drag forces from the small lubrication gaps between neighboring particles.  This suggests a critical shear rate above which particles stick together transiently by the lubrication forces and can grow into larger clusters.  At lower shear rates, the particles motions are more independent.  The large clusters result in a larger effective viscosity, so this critical shear rate $\dot\gamma_{min}$ and a corresponding shear stress $\tau_{min}$ signal the onset of shear thickening.  Hydroclustering is covered more thoroughly in a recent review \citep{WB09}.  This mechanism has produced viscosity curves in quantitative agreement with measurements for continuous shear thickening, in terms of both the critical shear rate and the magnitude of the increase in viscosity \citep{BBV02, MB04a,NM12}.  Furthermore, recent confocal rheology measurements have made it possible to directly observe clusters coinciding with the shear thickening regime, shown in Fig.~\ref{fig:hydroclusters} \citep{CMIC11}.  

 
It is often assumed that this model for shear thickening will produce discontinuous shear thickening at higher packing fractions, as particle clusters get larger and potentially span the system.  However, calculations and simulations based on the hydrocluster model so far have not been able to produce a viscosity increase greater than about a factor of 2, far less than the orders-of-magnitude increases in viscosity observed in experiments.  A key difficulty is that both calculations and simulations become increasingly more difficult at higher packing fraction and therefore have not been done close to the jamming transition.  Thus, it remains an open question whether  hydroclustering can lead to the steep viscosity curves that are the defining feature of DST.

The major quantitative tests of the hydrocluster model so far have focused on the onset stress for shear thickening, $\tau_{min}$. Since this has agreed with experiments for a wide variety of both continuous and discontinuous shear thickening systems, it has been widely interpreted as evidence for the general validity of the hydrocluster model.  However, a deeper investigation unveils that all of the different models for shear thickening predict the same onset stress scales.  Understanding the significance of this requires a discussion in the context of the other mechanisms, to which we will return to in Sec.~\ref{sec:onsetstress}.  


\subsection{Order-disorder transition}

The order-disorder transition mechanism was first identified and developed by Hoffmann \citep{Ho74,Ho82}.  He found that in some cases a transition to DST coincides with a transition in the microstructure from ordered layers at lower shear rates to a disordered state at higher shear rates.   Like the hydrocluster model, this scenario has been successful at predicting the onset shear rate $\dot\gamma_{min}$.

However, it has been  shown definitively that DST can occur without  an order-disorder transition \citep{MW02, EW05, ENW06}.  Thus, while the order-disorder transition is a possible way in which a microstructural reorganization coincides with shear thickening, it is not a required mechanism.

\subsection{Dilatancy}

It has long been known that dilatancy is  observed along with DST \citep{Re1885, MW58}.   Dilatancy also is a feature of dense granular flows in which, when sheared, the particles try to go around each other but often cannot take a direct path, so their packing volume expands (dilates) \citep{Re1885,OL90}.  In fact, in some of the early literature `dilatancy' was used as a synonym for shear thickening \citep{FR38, MW58,Ba89}, but this fell out of style after a landmark paper by Metzner and Whitlock \citep{MW58}. They confirmed that in many cases the onset of dilatancy and DST coincided, but that in some suspensions dilatancy could be observed without shear thickening. Since this ruled out a one-to-one correspondence, it was concluded in the rheology community that dilatancy should not be considered directly related to shear thickening.  For about 40 years following this result, many of the major papers on shear thickening dropped the connection to dilation in favor of hydrodynamic descriptions \citep{Ho82, BB85}.   However, there is another possible interpretation of the work of Metzner and Whitlock \citep{BJ12}.  Inductively, their observations suggested that dilation was necessary but not sufficient for DST.  Recent work has identified additional conditions that can explain why DST is not always observed along with dilation: in many suspensions shear thickening can be hidden by a yield stress or other source of shear thinning behavior \citep{BFOZMBDJ10}.  




 Much recent work  has led to a reconsideration of dilatancy as a mechanism for shear thickening and culminated in the development of a model that explains how dilatancy can lead to DST \citep{OM00, HFC03, LDH03, LDHH05, CHH05, FHBOB08, BJ12}.  The basic idea is that when dilation of granular shear flows is frustrated by boundary conditions that confine the suspension, shear results in normal stresses against the boundaries.  Confinement can provide an equal and opposite restoring force which is transmitted along frictional (solid-solid) contacts between neighboring particles that participate in a fabric of force chains.  The frictional contacts produce shear stresses proportional to normal stresses, enabling the dramatic increase in shear stress with shear rate associated with DST \citep{BJ12}.

This model has been developed to the point of quantitatively predicting both the lower and upper stress bounds on the shear thickening regime, $\tau_{min}$ and $\tau_{max}$, discussed in detail in Sec.~\ref{sec:stressscales}.  Like the other models, it still does not quantitatively predict the slope of the viscosity curves.  But it is able to explain the normal force measurements observed in DST systems, and an unusual dependence on the boundary conditions, which will be discussed in Sec.~\ref{sec:normalstress}.

\subsection{Equation of state models}

There are several phenomenological models of shear thickening that use equations of state to establish the relationship between stress and shear rate.  For example,  Cates and coworkers have used a scalar differential equation with an effective temperature to drive the dynamics  \citep{HAC01, HFC03, HCFS05}.  To produce shear thickening, this model requires an ad hoc assumption that the effective temperature decreases with increasing stress.  Another model uses  a microstructural state variable that  varies with stress, which can lead to shear thickening if this variable is assumed to have a critical point \citep{NNM12}.  These models have made several  predictions such as hysteresis in $\tau(\dot\gamma)$ such that the critical shear rate or stress for the onset of shear thickening differs depending on the direction of the time derivative of the shear rate or stress \citep{HAC01}.  This hysteresis is commonly observed in measurements of shear thickening \citep{CHH05, De10}.  A further interesting prediction has been the occurrence of  oscillations between high and low branches of the viscosity curve \citep{AC06, NNM12}, but this has not yet been observed in shear thickening suspensions.  

\section{Stress scales}
\label{sec:stressscales}

\subsection{Onset stress}
\label{sec:onsetstress}

The major quantitative test and success of both the hydrocluster and order-disorder transition models has been the prediction of the onset shear rate $\dot\gamma_{min}$.  Experiments have revealed that the onset shear rate varies with suspension viscosity -- which depends on packing fraction as well as the suspending liquid viscosity -- such that the onset of shear thickening is more simply characterized by an onset stress $\tau_{min} = \eta\dot\gamma_{min}$ that is roughly independent of packing fraction and liquid viscosity \citep{FHBM96, MW01a, SW05, BJ09}.  Thus, it is simpler to discuss the scaling of the onset of shear thickening in terms of this stress scale $\tau_{min}$, which then can be more directly related to the different forces between neighboring particles in densely packed suspensions.  We will further argue that such a description of shear thickening in terms of stress scales does not only enable a distinction between different mechanisms of stress transfer, i.e. viscous drag vs. solid-solid friction between particles, but also allows for a more general description of shear thickening mechanisms that  apply across a range of parameters regimes, each with different dominant stresses.




 In early models for Brownian colloids the onset of DST was described by a critical Peclet number $Pe = 6\pi\eta_l\dot\gamma a^3/k_{B}T$ for a particle size $a$, liquid viscosity $\eta_l$, and thermal energy $k_{B}T$.  Shear thickening was expected to occur for $Pe \stackrel{>}{_\sim} 100$ as the shear stress overcomes thermal diffusion of the particles \citep{FMB97, BBV02}.  This model has been successful at calculating the onset of shear thickening for both continuous shear thickening and DST  when written in terms of a stress scale $\tau_{min} = 50k_{B}T/3\pi a^3$, independent of the liquid viscosity \citep{GZ04, MW01a}.  
 
 For colloids where repulsions from a zeta potential are dominant the above model had to be modified \citep{MW01a}.  In that case, the particular scaling found was a stress characterizing electrostatic repulsions between neighboring particles.  While the forces were calculated at a distance corresponding to an effective hydrodynamic radius obtained from a force balance between viscous and electrostatic forces, calculating at a different radius would only have changed the result by a scale factor of order 1.   Since the model was an order-of-magnitude calculation, it would have resulted in just as good a match with the data if a different radius was used.  In the end, the modifications to the hydrodynamic model required to fit it to the data resulted in completely eliminating any dependence on hydrodynamic parameters such as viscosity or shear rate.  The associated stress scale is thus not specific to hydrodynamic mechanisms for stress transfer, as any type of forces transferred through a continuum system can be expressed in terms of a stress.  
 
 With several relevant forces in colloids and suspensions, each of which could be dominant in different cases, a variety of different scalings for the onset have been found.  The common trend is that the onset can be described more simply in terms of a stress scale (rather than a shear rate) independent of packing fraction and set by some dominant force in the system.  Depending on the parameter range, this dominant force could be Brownian motion \citep{BBV02}, zeta potential \citep{MW01a}, particle-liquid surface tension \citep{BFOZMBDJ10}, induced dipole attractions \citep{BFOZMBDJ10}, or steric repulsion \citep{Ho98}.  Notably, in each case hydrodynamic terms such as shear rate and viscosity were absent from  the modified scalings which match the experiments, so this suggests inertia or hydrodynamics-based models are not necessary to determine the onset of DST as initially envisioned by the Peclet number scalings.   In all cases, the onset of shear thickening has been rationalized in terms of a stress scale $\tau_{min}$, although the value of $\tau_{min}$ depends on a dominant stress scale  of the system.

Here we discuss the onset stress scalings in the most general terms possible.  The common feature of the aforementioned scalings for the onset stress $\tau_{min}$ is that the shear stress must exceed all local stress barriers that are responsible for preventing relative shear between particles.   The significance of the shear stresses exceeding stresses that prevent shear is that  local shearing between grains can lead to dynamic particle contacts, dilation, increased confining stresses, and frictional stresses observed as shear thickening.  In the simplest cases, these stress barriers can come from particle attractions from various sources, including particle-liquid surface tension and induced attractions from external fields \citep{BFOZMBDJ10}.  In each of these cases the attractions resulted in a yield stress on the same scale as $\tau_{min}$ due to the attractions. The scale of $\tau_{min}$ was set by the shear stress required to overcome roughly the two-particle attractive force (per cross-sectional area of a particle) to shear them apart.


This picture can also apply to colloids with a repulsive electrostatic potential.    While attractive particles may have to be pulled apart to shear, repulsive particles may have to be pushed around each other to shear.  If the particles push against each other they end up pushing against all of the confining stresses, whose net response is still determined by the softest component of the system.  This means we expect the onset stress to be set by the scale of the two-particle interaction stress scale regardless of whether it is attractive or repulsive.   The similar behavior for both attractive and repulsive particles is analogous to jammed systems \citep{OSLN03}.  For colloids stabilized by an electrostatic zeta potential $\zeta$, the observed scaling for $\tau_{min}$ is proportional to the electrostatic repulsive force per cross-sectional area of a particle $16\epsilon\zeta^2/a^2$ for a liquid permittivity $\epsilon$ \citep{Ho98}.  This scaling is also consistent with predictions which were based on hydrodynamic models up to a dimensionless coefficient of order 1 \citep{Ho82, MW01a}.

In suspensions of particles large enough to settle the scale of $\tau_{min}$ is set by gravity rather than attractions \citep{BJ12}.  The shear stress needs to be enough to exceed the weight of a particle per cross-sectional area to overcome friction.  This follows the rule of dominant stress scales, despite the fact that such suspensions are inhomogeneous.

In the Brownian-motion dominated regime, the onset stress $\tau_{min}$, as shown above, 
corresponds is the osmotic pressure, which is an effective repulsive stress between neighboring particles.  Again, this scaling was originally derived from hydrodynamics-based models, but again the hydrodynamic terms cancel out.  It is notable that this scaling for the onset works both for continuous shear thickening and DST \citep{MW01b,GZ04}.  The generality of the onset scalings arises because the scalings for $\tau_{min}$ are set by mechanisms for shear thinning which are independent of the mechanisms for shear thickening.   Either type of shear thickening can be hidden until the stresses from shear thickening mechanisms exceed all stresses from shear thinning mechanisms \citep{BFOZMBDJ10}.  This argument is simply based on which stresses are dominant, so it is not specific to a particular mechanism or to whether the shear thickening is discontinuous or continuous.

The fact that the onset stress can be described by such a general argument in terms of the dominant stress scales without the need to specify microstructure or the mechanism of force transfer means that the onset stress cannot be used to distinguish between the different mechanisms proposed for shear thickening.  In particular, the quantitative models (hydroclustering, order-disorder transition, and dilatancy) all predict the same stress scales for the onset of shear thickening.  All three mechanisms have been observed in different cases along with shear thickening, and so it seems that they are all valid, albeit not unique, microstructural mechanisms for triggering its onset.

\section{Coupling of normal and shear stress}
\label{sec:normalstress}

\begin{figure*}                                         
\centerline{\includegraphics[width=7.in]{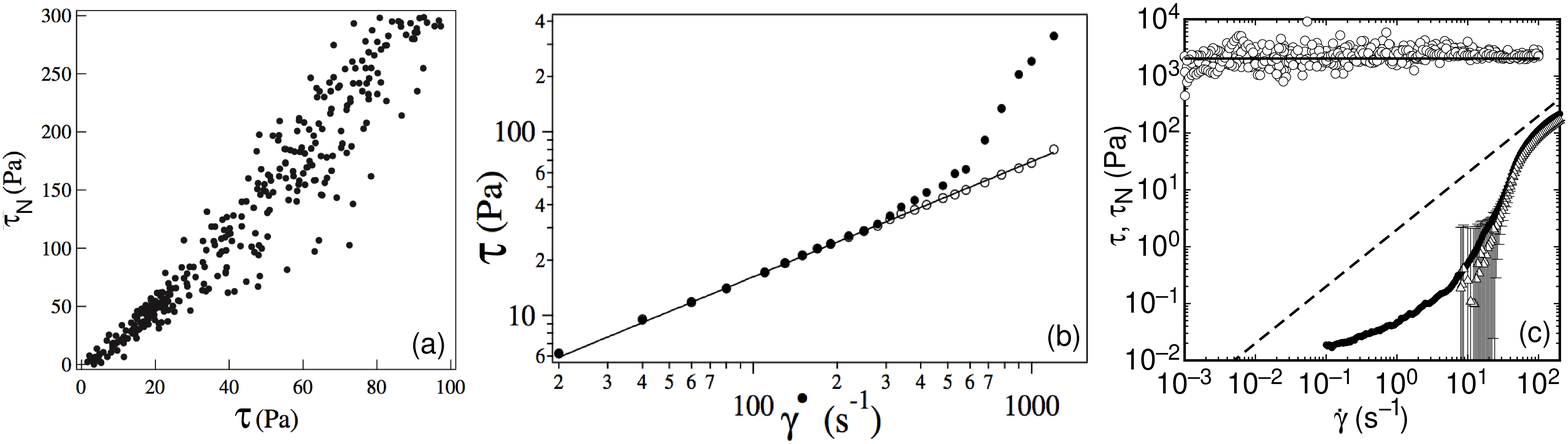}}
\caption{ Coupling of the normal and shear stresses in DST. (a) Linear proportionality between shear stress $\tau$ and normal stress $\tau_N$ in the giant fluctuations of a shear thickening fluid measured at constant shear rate.  Panel reproduced from \citet{LDHH05}.  (b)  Viscosity curve for a shear thickening fluid, based on the average shear stress (solid circles) and the most probable shear stress (open circles) corresponding to a baseline stress value without giant fluctuations. The baseline has a Newtonian scaling, suggesting the shear thickening is due to the giant fluctuations, which are strongly correlated to the normal stress. Panel reproduced from \citet{LDH03}.  (c) Comparison of flow curves measured with different boundary conditions. Solid circles: shear stress $\tau$ from a fixed-gap measurement with a standard parallel plate setup. Open triangles: normal stress $\tau_N$ from the same measurement. The absolute uncertainty on the normal stress is 2 Pa, so the normal stress cannot be resolved at the low end. Open circles: $\tau$ for fixed normal stress of 2040 Pa (solid line) in a modified parallel plate setup with a hard wall. In both experiments, the shear and normal stresses are strongly coupled, but the relationship between stress and shear rate changes dramatically with the boundary conditions: in one case a yield stress fluid results and in the other DST.   Panel reproduced from \citet{BJ12}.
} 
\label{fig:normalstress}                                        
\end{figure*}


In this section we discuss a number of observations that demonstrate a strong coupling between shear and normal stresses.
DST systems can exhibit remarkably large fluctuations.  For example, in steady state rheometer measurements under constant applied shear rate, time series of the shear stress can  fluctuate  more than an order-of-magnitude in the shear thickening regime \citep{LDH03}.  These fluctutations in the shear stress are  strongly correlated to the normal stress, with a direct proportionality between shear and normal stress (proportionality factor of order 1) \citep{LDHH05}.  A  plot of $\tau$ vs. normal stress $\tau_N$ for such fluctuations is shown in Fig.~\ref{fig:normalstress}a.  These fluctuations in shear stress are largely positive on top of a baseline Newtonian $\tau(\dot\gamma)$ \citep{LDH03}.  This is demonstrated in Fig.~\ref{fig:normalstress}b.  It compares $\tau(\dot\gamma)$ as obtained from an average over an entire time series, which exhibits shear thickening,  with the mode stress value at the baseline, which scales as a Newtonian fluid.   These observations suggest that without the fluctuations, these suspensions are Newtonian and the baseline stresses are mainly viscous in nature.  DST, then, is a result of fluctuations which are coupled to the normal stress.

The coupling between shear and normal stresses is so strong that it supersedes an intrinsic $\tau(\dot\gamma)$ relation and survives with different boundary conditions.  As an example, the shear stress $\tau$ and normal stress $\tau_N$ are shown for a DST suspension in Fig.~\ref{fig:normalstress}c as function of shear rate $\dot\gamma$ for a typical rheometer measurement in which the gap size has been fixed.   Positive normal stresses are generally found, corresponding to the sample pushing against the plates, again strongly coupled to shear stresses \citep{JR93, LDHH05, FHBOB08, BJ12}.  This is compared with an experiment in a similar suspension in which the normal force has been fixed and the gap size can vary as required.  While there is still a strong coupling of the shear stress to the normal stress, now the rheological behavior is that of a yield stress fluid with no shear thickening regime \citep{BJ12}.  Such dramatic difference in behavior with change in boundary conditions suggests that DST is not an intrinsic bulk property of the fluid.  A variety of different experiments have similarly shown a strong coupling between shear and normal stresses and a violation of the assumption of a direct intrinsic relation between shear stress and shear rate \citep{LDHH05, FHBOB08, BJ12}.


This coupling between $\tau$ and $\tau_N$ implies a redirection of stress by particle interactions within the bulk of the suspension  \citep{NB94, PK95, BV95, JNB96, SB02, DGMYM09}.   Since there is not an intrinsic relationship between stress and shear rate, and the coupling between stresses is observed even during transients and under different boundary conditions,  this supports the idea that the stresses do not come from viscous lubrication but instead from solid-solid friction in which  forces are transmitted along chains of hard particles via frictional contacts \citep{JNB96}.

\subsection{Limits on lubrication}



Since most early models for shear thickening were based on hydrodynamics such that the effective viscosity is dominated by the flow in the lubrication gap between particles \citep{BB85}, it is instructive to highlight a relevant limitation of lubrication models.  By relating the lubrication drag force between particles to the size of the gap between particles, one can put a strict upper bound on the effective viscosities reached due to viscous drag forces between particles.

From lubrication theory, the effective viscosity can be estimated assuming laminar squeeze flow of liquid between neighboring particles of diameter $a$ in suspension spaced apart by a characteristic gap size $h$ \citep{FA67}.  The effective viscosity is $\eta = C\eta_l (a/h)$ where $\eta_l$ is the liquid viscosity and $C$ a geometric constant of order 1.  The above acts as a way to estimate the gap size required to obtain a particular viscosity scale.   If the gap becomes as small as two molecular layers, the continuum fluid model breaks down and the molecules interact mechanically as if they are frictional solids \citep{VG88}.  Therefore, the maximum effective viscosity due to viscous lubrication is bounded by setting $h$ to two molecular layers.   In real flows, the particle gaps are not uniform, so some particles would undergo solid-solid contact friction earlier than this bound would indicate.

As an example, typical cornstarch particles have an average diameter of 14 $\mu$m, so the upper bound on the lubrication contribution to the viscosity for a suspension of cornstarch in water is $4\times10^4$ times the viscosity of water when the lubrication gaps becomes 2 water molecules thick (Using $C=9/4$ \citep{FA67}).   However, suspensions of cornstarch in water have been measured to have effective viscosities up to at least $10^7$ times the viscosity of water in the shear thickening regime \citep{BJ09}.  This is orders of magnitude greater than possible with lubrication contributions to the viscosity. This also means that lubrication forces cannot generally support the large stresses observed during DST, because the gaps between particles would reach 2 molecular layers and the particles would interact as if they have solid-solid contacts before the highest stresses are reached.  Nevertheless, models based on hydrodynamic interactions such as the hydrocluster model can still be valid in the parameter range relevant to the onset of DST or during continuous shear thickening, where the effective viscosity is still low enough that lubrication forces may  dominate.


\begin{figure}                                                
\centerline{\includegraphics[width=3.4in]{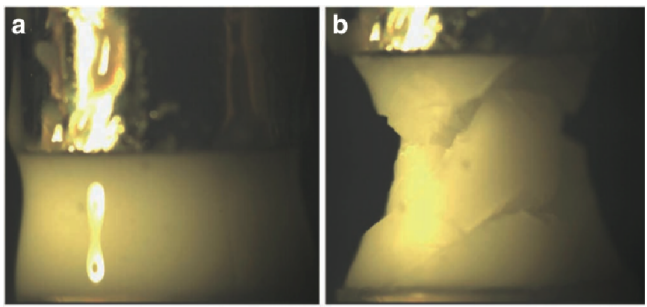}}
\caption{During shear, dilation causes particles to poke through the surface as the interstitial liquid  retreats into the center, giving the surface a rough appearance.  Here, exemplified by side views of a suspension  of 0.6 $\mu$m diameter PMMA spheres in stearic acid between two metal plates  (a) at rest and (b) during vertical extension that stretches the fluid.  At rest, the surface is shiny.    Such visible change in the surface portends an important change in the boundary condition, where surface tension can produce strong forces on particles and keep them jammed.  Figure reproduced from \citet{SBCB10}.
} 
\label{fig:surfacedilation}                                        
\end{figure}

When lubrication breaks down at high stresses, the frictional coupling between shear and normal stresses is provided by disordered, dynamically reconfiguring structures (force chains) that extend all the way to the boundaries.  These structures may arise as a result of hydroclusters growing in size to span the system \citep{WB09}.  While the growth and evolution of this frictional contact network  has  not yet been established via direct experimental observation  in dense suspensions, there are many observations of the consequences, in particular the dilation of the particle packing along with DST \citep{FR38, MW58, OM00, FHBOB08, BJ12}.  A further realization over the last decade has been that dilatancy leads to an important role of capillary forces at boundaries \citep{HFC03,HCFS05, CAS05, BJ12}.  

A number of  experiments have shown that when densely packed suspensions of particles are subjected to any type of shear, the concomitant dilation requires an increase in available volume in the bulk and results in the interstitial liquid retreating to the interior so that the particles appear to poke out of the surface \citep{CHH05, SBCB10, KW11, BJ12, MJ12}.  If the particles are between about 1 to 100 microns in size, the surface appears by eye to change from shiny to rough as a result of this dilation, as the partially exposed particles  scatter light diffusely.  An example of this surface change can be seen in Fig.~\ref{fig:surfacedilation} for a suspension under tensile stress.

Dilation introduces additional stresses on the suspension due to interaction with the boundaries which confine the suspension \citep{Re1885, OL90}.   In a typical suspension that is open to the air, the liquid-air surface tension at the boundary provides a force that pushes penetrating particles toward the interior.   This is the force that holds the suspension together so particles do not fall out of the liquid.  For the particles to penetrate the surface in a steady state, as observed during DST, these forces must be transmitted through the interior along force chains.  The frictional contacts and redirection of forces along these chains requires the strong coupling between shear and normal stresses to obtain a net force balance in the steady state.  Thus, dilation is a mechanism by which normal and shear stresses become coupled in a frictional relationship, and the suspension rheology becomes dependent on the boundary conditions.

 \subsection{Maximum stress scaling with the boundary stiffness}
\label{sec:taumax}

\begin{figure}                                                
\centerline{\includegraphics[width=3.4in]{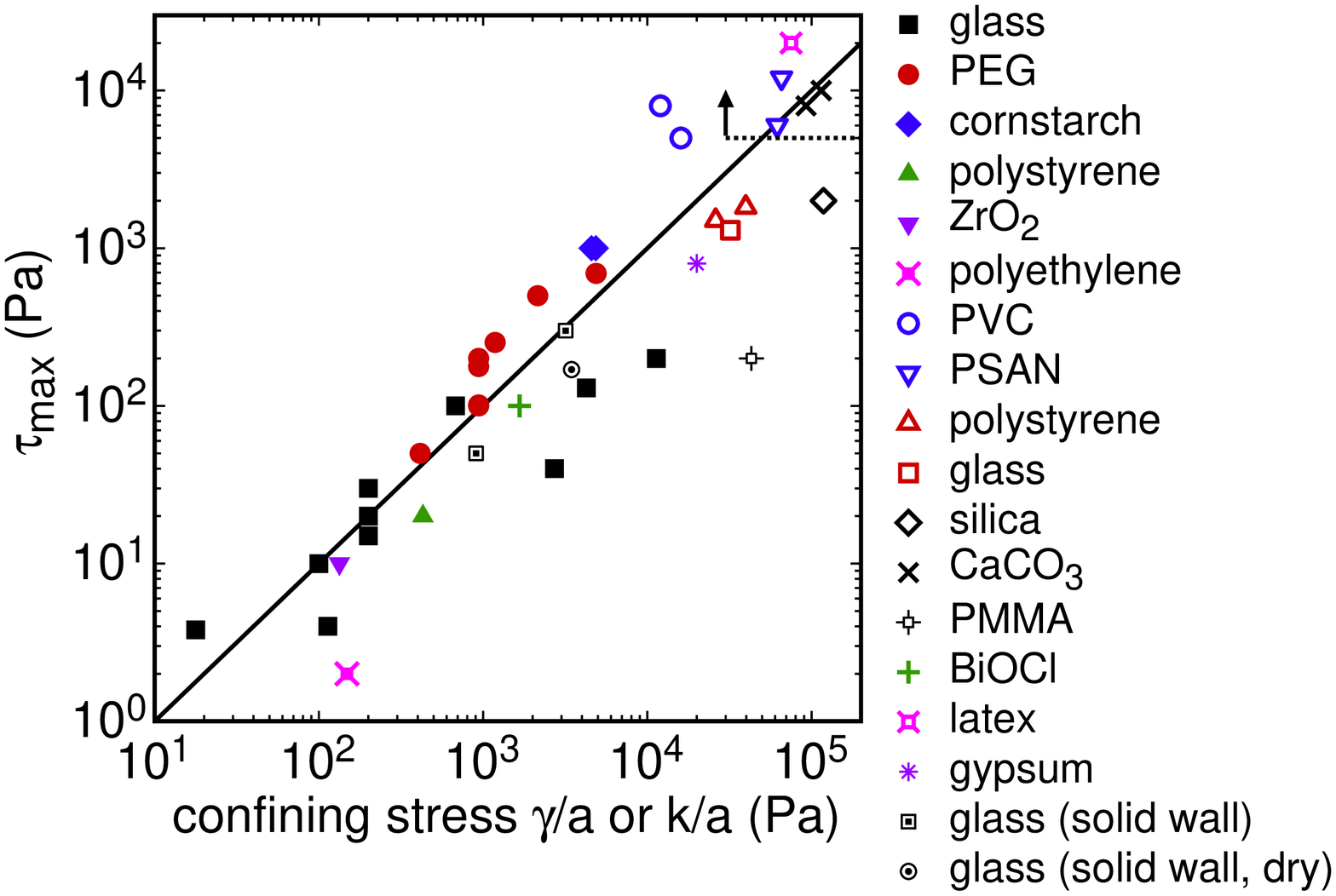}}
\caption{Stress at the upper boundary of the shear thickening regime, $\tau_{max}$, plotted against the confining stress scale from surface tension $\gamma/a$ for a variety of suspensions.  Particle materials are listed in the key.  Solid symbols: \citep{BJ12}.  Open symbols: polyvinyl chloride [PVC, circles \citep{Ho72}], polystyrene-acrylonite [PSAN, down-pointing triangles \citep{Ho72}], polystyrene [up-pointing triangles \citep{BBLS91}], glass [square \citep{BLS90}], silica [diamond \citep{BW96}], CaCO$_3$ [diagonal crosses \citep{EW05}], PMMA [crossed square \citep{KMMW09}], BiOCl [cross \citep{BBS02}], latex [diagonally crossed square \citep{LBS91}], gypsum [8-point star \citep{NBPVBM12}].  The solid line corresponds to a scaling $\tau_{max} = 0.1\gamma/a$.  Dotted line: lower bound on $\tau_{max}$ for measurements in which $\tau_{max}$ was not reached \citep{MW01a}, which often occurs in colloid measurements due to limitations of rheometers.  Also shown are results from experiments with solid walls at all boundaries, using glass spheres suspended in water (partially filled squares) or without interstitial liquid (partially filled circle). In these cases the confining stress scale is replaced by $k/a$, where $k$ is the effective stiffness per unit particle cross-section of the wall.  Figure based on \citet{BJ12}.
} 
\label{fig:stressmaxcollection}                                        
\end{figure}

The knowledge that dilation couples the shear and normal stresses to the boundary allows a prediction of how strong shear thickening can become, i.e., a prediction of the scale of $\tau_{max}$. In a typical suspension open to the air, for example around the perimeter of a rheometer tool, the liquid-air surface tension $\gamma$ at that boundary provides a force that pushes penetrating particles toward the interior with a stress that is on the order of $\gamma/r$, where  $r$ is the radius of curvature of the liquid-air interface.  Without deformation of the interface by particles, $r$ would be determined by the tool and container geometry or the capillary length.  But in a dense suspension that dilates under shear, the scale of $r$ decreases as particles deform the interface until it is limited geometrically by the scale of the particle diameter $a$.  Thus, the confining stress at the suspension-air interface is on the order $\gamma/a$, much larger than in a Newtonian liquid \citep{LNS02, HFC03, MB04b,HCFS05, BZFMBDJ11, BJ12}.  This implies that normal and shear stresses are limited by the confining stress scale $\gamma/a$ from the boundary.  In rheological measurements this limiting strength corresponds to the upper end of the shear thickening regime $\tau_{max}$.  Beyond $\tau_{max}$, any additional shear stress must come from other sources, which are likely weak compared to the confining stress if shear thickening is observed, so the viscosity decreases beyond $\tau_{max}$.


Measured values of $\tau_{max}$ are plotted vs. the confining stress $\gamma/a$ in Fig.~\ref{fig:stressmaxcollection}.  Each point corresponds to a different DST suspension, with a wide range of different particle materials, shapes and sizes, and different liquids.    It is seen that for this wide variety of suspensions, covering four decades, $\tau_{max}$ falls in a band with a scaling proportional to $\gamma/a$.  While most measurements are done in shear flows,  slightly larger values of $\tau_{max}$ are obtained for extensional flows \citep{WCR10, SBCB10}.   In many measurements of colloids, the upper end on the shear thickening regime was not reached.  This is especially a problem in the colloid regime because the expected scale of $\tau_{max}$ for small particles exceeds the measuring range of many rheometers.   A lower bound on $\tau_{max}$ based on the limited measuring range is illustrated as the dotted line in Fig.~\ref{fig:stressmaxcollection}, using data from \citet{MW01a} as an example.



 Within the band shown in Fig.~\ref{fig:stressmaxcollection} there is variation by about an order of magnitude in the value of $\tau_{max}$. This is likely due to a number of dimensionless  factors of order 1 that  contribute to the precise value of the confining stress and the resulting shear stress.  These include the effective coefficient of friction  \citep{Janssen1895, Sperl06}, the contact angle for wetting, as well as factors related to particle geometry.  Nonetheless, the scaling $\tau_{max} \sim \gamma/a$ is found to hold as an approximate scaling for a wide range of suspensions.


Observations have also been made of DST in suspensions contained by solid-wall boundaries, without any suspension-air interface.   Notably, the shear thickening is similar with and without any interstitial liquid \citep{BJ12}.  This unambiguously demonstrates that DST can occur without any lubrication forces between particles.  Some data is included in Fig.~\ref{fig:stressmaxcollection} for experiments with solid walls at all boundaries (partially filled symbols).  In this case the stress scale is $k/a$, where $k$ is the stiffness of the boundary over an area with cross-section equal to that of a particle, analogous to a surface tension.  The fact that the scaling of $\tau_{max}$ is similar to that when the boundary stiffness is determined by surface tension suggests that $\tau_{max}$ is generally determined by a confining stress at the boundary, irrespective of source.  

When a suspension is confined by  boundaries of differing stiffness, it should be understood that the least stiff boundary limits $\tau_{max}$.  Since forces are easily redirected throughout the suspension, every boundary must be able to support $\tau_{max}$ in a force balanced steady state. This is analogous to a system of elastic materials in series, where the least stiff material controls the overall system stiffness.


\subsection{State diagram}

 \begin{figure}
\centerline{\includegraphics[width=3.2in]{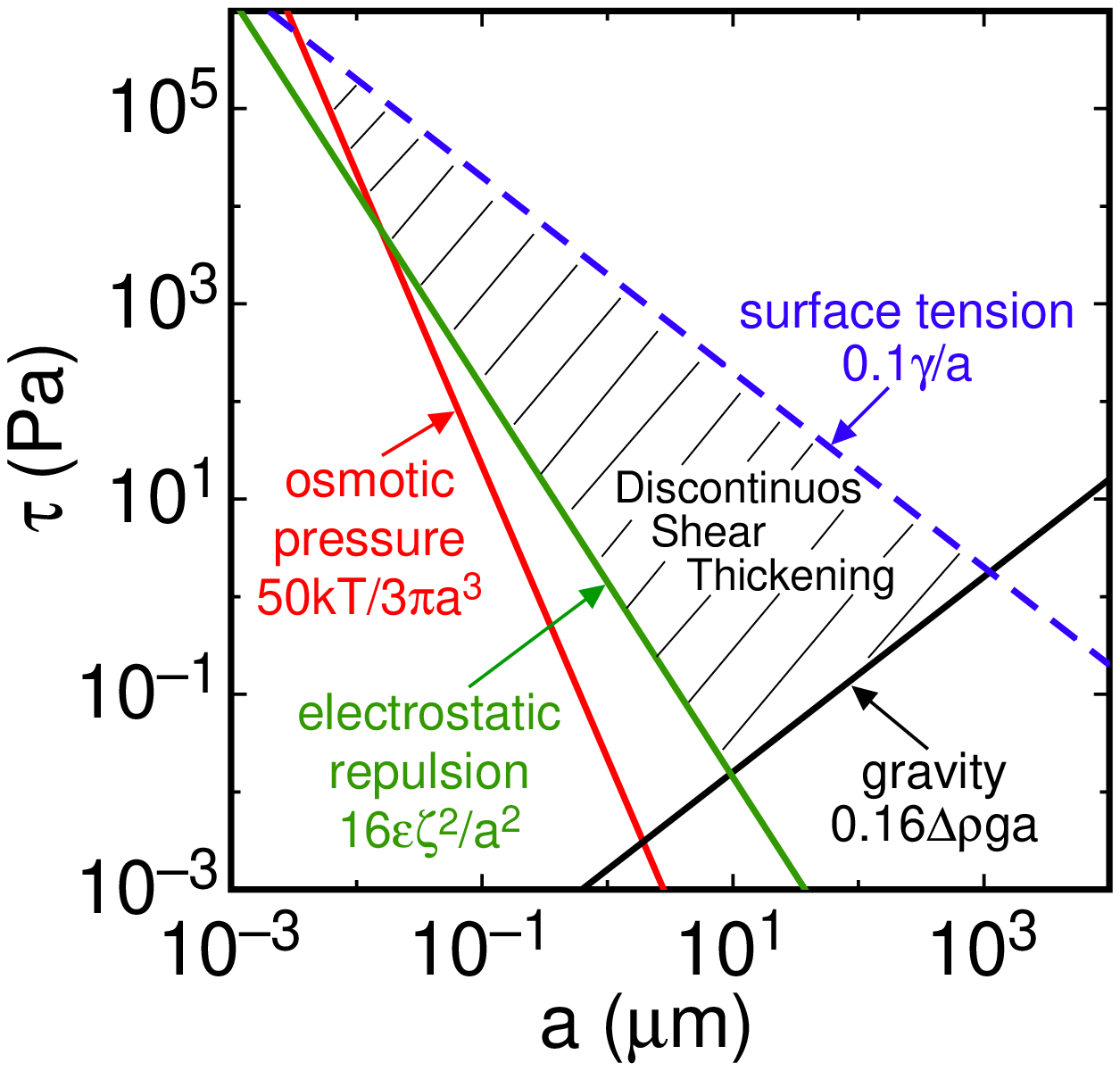}}
\caption{Rheological state diagram for a hypothetical suspension including all known scaling laws for shear thickening regime boundaries.  It is assumed the suspension has a liquid-air interface at the boundary and the liquid wets the particles. Red solid line: osmotic pressure at room temperature \citep{MW01b, GZ04}.  Green solid line:  Electrostatic repulsion for a surface potential $\zeta=70$ mV  \citep{MW01a, Ho82}. Black solid line: gravitational scaling with a density mismatch $\Delta\rho = 1$ g/mL \citep{BJ12}.  Blue dashed line: confining stress from surface tension with $\gamma = 20$ mN/m \citep{BJ12}.     Since each boundary depends on different parameters, each can be tuned independently; in many suspensions a shear thickening regime may not occur if the scalings for the confining stresses fall below the scalings for stresses that oppose particulate shear.
}
\label{fig:phasediagram}                                        
\end{figure}
 
 The boundaries of the shear thickening regime, $\tau_{min}$ and $\tau_{max}$, reflect the dominant stress scales  that oppose shear between neighboring particles and  that provide a confining stress in response to dilation, respectively.   In this section we develop these ideas into a state diagram.  To delineate a typical parameter regime for DST in suspensions, we show how each of the  scalings mentioned earlier provide bounds for the shear thickening regime in Fig.~\ref{fig:phasediagram}.  Since no single suspension material covers the entire parameter space, we give scalings for a hypothetical suspension with some typical material properties, but note that each of the boundaries can be tuned independently depending on particle and liquid parameters.

As can be seen in Fig.~\ref{fig:phasediagram}, $\tau_{min}$ for osmotic pressure and electrostatic interactions tend to be dominant for smaller particles in the colloid regime, while the stress scale for gravity is dominant for larger particles in the suspension regime.  These scalings for $\tau_{min}$ typically meet at a size of around 10 $\mu$m depending on the values of zeta potential, density, and so on.  This suggests that suspensions of particles on this size scale will tend to have the smallest values of $\tau_{min}$.   The minimization of $\tau_{min}$ defines an optimal particle size for shear thickening, where the largest (logarithmic) stress range for the shear thickening regime will typically be found.   Physically, this optimal size corresponds to the colloid-suspension transition, which is effectively defined by the transition between the dominance of Brownian motion and electrostatic repulsion in the colloid regime to the dominance of gravity in the suspension regime.



The maximum particle size at which shear thickening was found is about 1 mm \citep{BJ12}.  An upper bound is expected when $\tau_{min} \sim  \Delta\rho g a$, which is set by gravity, and increases with particle size, meets with $\tau_{max} \sim \gamma/a$, which is set by surface tension, and decreases with particle size, as seen in Fig.~\ref{fig:phasediagram}.  This balance corresponds to a particle capillary length scale $a \sim \sqrt{\gamma/(\Delta\rho g)}$ which differs from the usual capillary length in two ways.   First, this particle capillary length depends on the density difference $\delta\rho$ rather than just the liquid density.  Second, this particle capillary length sets a transition between scaling regimes based on particle size rather than system size.  This means surface tension effects can be seen in suspensions on much larger scales than would usually be expected based on the usual capillary length.  

The particle capillary length should typically be around 1000 $\mu$m for most suspensions, in agreement with the maximum size particle found to shear thicken \citep{BJ12}.  While this gives a typical particle size scale above which most particles will not shear thicken in suspension, it is less tied to this size scale than the macroscopic capillary length because it depends on the density difference. Thus, the maximum particle size could in principle be much higher for carefully density matched suspensions.

While we have described a mechanism for shear thickening that is based on generic phenomena such as dilation, not all suspensions and colloids are reported to shear thicken.  This can  be explained partly by the relative importance of different stress scales.  If any other particle interaction scales exceed the confining stress from surface tension, we would expect shear thinning mechanisms to be dominant over shear thickening \citep{BFOZMBDJ10}.  In terms of the state diagram, if the lower bound of the shear thickening regime $\tau_{min}$ exceeds the upper bound $\tau_{max}$, then there is no shear thickening regime in between.  This situation is quite common for real suspensions in the colloidal regime.  We have also left some such particle interactions out of the state diagram because the corresponding scaling laws for $\tau_{min}$ are not as well established.  These interactions include hydrogen bonding \citep{RWK00}, depletion \citep{GZ04}, or a particle-liquid surface tension \citep{Ba89, BFOZMBDJ10}.  Thus, the rarity of observations of shear thickening in dense suspensions and colloids can in part be explained by the fact that many colloids fall into the regime where the stress scale of particle interactions exceeds the confining stress scale so they do not have any shear thickening regime.  Another likely reason for the apparent rarity of shear thickening is that it occurs in a fairly small parameter space with a narrow range in packing fraction, so many measurements of suspension rheology simply do not cover this range.

It is provocative that cornstarch, arguably the most famous shear thickening particle, is on the optimal size scale of around 10 $\mu$m.  In terms of chemical and physical properties,  it is notable only in that it is extremely hygroscopic. This implies minimal particle-liquid  surface tension and consequent shear thinning effects in water \citep{BFOZMBDJ10}, which also happens to have one of the highest surface tensions of common liquids.  Cornstarch remains an inert, hard particle (with Young's modulus around ~10 GPa \citep{JHMK13}), in contrast to some other mass-produced powders such as flour, which gels in water at room temperature.  Thus we attribute the  strong shear thickening of cornstarch to its optimal particle size and lack of the various interactions which produce shear thinning effects that could hide shear thickening.



 \section{Relation to Jamming and other soft matter systems}
 \label{sec:jamming}

 \subsection{Broader view of physics of concentrated many-particle systems}
\label{sec:particulate}

Why DST is only found in the parameter regime of hard-particle suspensions and colloids, and not in emulsions, foams, or other dense suspensions of deformable particles  \citep{NVABZYGD10} was formally posed in Barnes' 1989 review and has remained a major question in the field.  It has been especially puzzling since models for shear thickening require only simple generic interactions and microstructural changes that could in principle occur in any type of complex fluid consisting of many particles.

To address this question, it is instructive to first consider various situations with hard particles.  Dry grains in an open container are not known to shear thicken.   When they shear they dilate but the free surface does not to provide an interface with a restoring force.  There is confining stress from gravity, but since it provides confinement even without shear it also sets the scale of the yield stress \citep{FBOB09}, and it does not produce a shear thickening regime.  It is only when when a confining stress is provided by enclosing the system with solid walls that shear thickening can be found for dry grains \citep{BJ12}.  This source of confining stress is not shown in Fig.~\ref{fig:phasediagram} because the scaling is not yet well established.  This observation also makes it clear that one of the important differences between dry and wet grains is that the surface tension of the liquid provides a confining stress.  

Some measurements of sheared dense suspensions in closed systems found inertial scaling rather than DST \citep{Ba54}.  While that system was enclosed, a rubber sheet was placed in between the suspension and the wall to allow dilation of the suspension, and a liquid reservoir allowed liquid to fill the gaps enlarged by dilation.  Thus, it seems likely that the rubber sheet was soft enough that its compression did not provide a significant confining stress in excess of the inertial contribution.  This suggests a possible method for greatly reducing the resistance in pipe flow of dense suspensions, namely to use compliant walls.

On the other hand, a closed system with very hard walls is expected to cause the grains to jam as there is no room for dilation and the hard walls would be able to apply enough stress to completely frustrate dilation.  This effect has been seen for hard disks  just below the onset of jamming based on uniform compression.   The disks contacted each other via force chains when sheared quasi-statically, i.e. the system jammed rather than shear thickened \citep{ZMB08}.   The yield stress in this jammed state with hard walls scales with the particle modulus as the particles compress against each other, which is the most compliant component of the system if the walls are harder than the particles.  This results in a different scaling for the yield stress with packing fraction than a suspensions with a liquid-air interface, since with hard walls the confining stress increases as the system is further compressed to higher packing fractions \citep{OSLN03}, while for a suspension the confining stress is limited by the scale $\gamma/a$ regardless of further compression.


The above situations with hard particles highlight the importance not only of dilation of the packing under shear, but also that the dilation must be partially frustrated by a restoring force from the boundary to produce DST.  This understanding allows us to address the case of complex fluids consisting of soft particles.  Foams and emulsions are prototypical systems used for jamming experiments with soft particles.   Rheologically they are generally found to shear thin, even in confined volumes.   In jammed foams, for example, the yield stress is observed to be on the scale of $\tau_y \approx 0.05\gamma/a$ \citep{GDJC98}.  This is the same relation we find for $\tau_{max}$ in response to dilation or for jammed suspensions due to the deformation of the liquid-air interface.  Since foam bubbles are very soft, they will be the limiting factor that determines the confining stress under almost any boundary conditions.  They are so easily deformable that they will typically shear without the need for dilation even in very dense packings.  Since this stress is very low, it seems unlikely that it can exceed attractive interactions considering they both come from surface tension.  As a result DST should not be expected in foams. Similarly, emulsions are very soft particles with stiffness set by (interfacial) surface tension, and so the confining stress in response to dilation is too small to expect DST. 


There is an intermediate regime where the particles are less stiff than the boundaries, but stiff enough that the confining stress would still exceed sources of $\tau_{min}$.   For example, DST has been observed in simulations of dry granular packings of elastic particles in periodic boundary conditions.  In this case, there is no hard boundary, so the confining stress comes from the stiffness of particles as they deform when dilation is frustrated \citep{OH10}.

These observations also help explain why the majority of simulations of suspension rheology based on lubrication theory have failed to produce DST, even though many have produced milder, continuous shear thickening \citep{MB04a,WB09,NM12}.  Most simulations remain focused on lubrication forces and the local microstructure.  Lubrication forces alone are not enough to produce DST, and some confining stress is needed.  The few simulations that have produced strong shear thickening include elastic forces between particles \citep{OH10,ZSX13}, so confining stresses could result from particle deformation.  In contrast, lubrication theory based simulations  tend not to account for particle deformation and stiffness, which can become impractical to model at high packing fractions simultaneously with lubrication hydrodynamics.  Thus, lubrication theory based simulations typically have not been operating in the right parameter regime to observe DST.  

 

\subsection{Critical point at the jamming transition}

DST was one of the phenomena first motivating the notion of jamming \citep{CWBC98} and DST has since often been connected  to jamming \citep{HAC01, MW01a, FHBOB08, BJ09}.   Jamming and  DST are both associated with a transition from a flowable to a solid-like state in a medium of randomly configured particles.  In both DST and jamming, forces are transmitted all the way across the system  along a fabric of local, solid-solid particle contacts.    The strength of DST systems is found to be limited by a confining stress ($\tau_{max}$), similar to the yield stress of a jammed system \citep{OSLN03}; in the latter case the confining stress is traditionally determined by the particle stiffness rather than surface tension as jamming is typically studied without interstitial liquids and with periodic boundary conditions.  In contrast with DST, in the original formulation of the jamming phase diagram, the jammed state is associated with a static, or at least not continually deforming, particle configuration that has not yet fully yielded to shear  \citep{LN98,OSLN03, LN10}.  There are more recent variations on this diagram in which the shear history can induce additional jammed configurations, but with an anisotropic fabric of stress-bearing contacts, at packing fractions slightly lower than the ordinary jamming phase transition \citep{MB05, BZCB11}. Similarly, DST occurs at packing fractions just below jamming, but emerges at finite shear rates, typically well beyond yielding, making it a dynamically driven state.  
While it may be useful to keep these differences in mind when dealing with specific circumstances, we nevertheless propose here that  the structural similarities warrant labeling the DST state a dynamically jammed state.

\begin{figure}                                                
\centerline{\includegraphics[width=3.in]{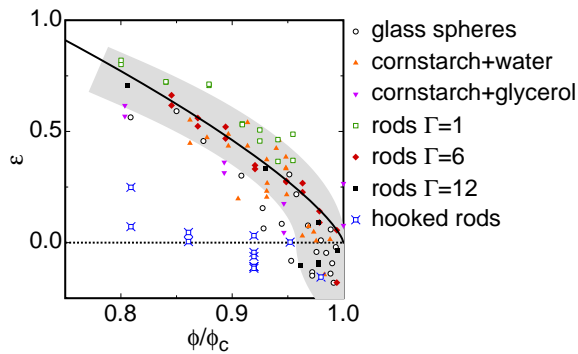}}
\caption{The strength of shear thickening is characterized by the exponent of a fit of $\tau\propto \dot\gamma^{1/\epsilon}$ to the shear thickening regime.  The inverse of the exponent $\epsilon$ is plotted vs.~packing fraction $\phi/\phi_c$ normalized by the jamming transition.  Data from several different suspensions including different particle shapes with very different values of $\phi_c$.  Black solid squares:  rods with aspect ratio $\Gamma=9$, $\phi_c=0.35$.  Red diamonds: rods with aspect ratio $\Gamma=6$, $\phi_c=0.37$.   Green open squares: rods with $\Gamma=1$, $\phi_c=0.55$.  Blue  crossed squares:  hooked rods, $\phi_c=0.34$.  Black open circles: glass spheres in water, $\phi_c=0.58$ \citep{BJ09}.  Purple down-pointing triangles: cornstarch in glycerol, $\phi_c=0.57$ \citep{BJ12}.  Orange up-pointing triangles: cornstarch in water, $\phi_c=0.48$ \citep{BJ12}.  The solid line is a best fit of $\epsilon \propto (\phi_c-\phi)^{\xi}$ to the data for convex shapes and $\phi/\phi_c > 0.8$.  The collapse of the data in the gray band suggests that the normalized packing fraction $\phi/\phi_c$ determines the strength of shear thickening for convex particle shapes.   Figure reproduced from \citep{BZFMBDJ11}.
}
\label{fig:epsilon_phirel}                                        
\end{figure}

We can relate DST to the jamming transition more quantitatively based on the divergent scaling of the viscosity curve.  In the shear thickening regime  $\tau(\dot\gamma)$ can be fit by  a power law $\tau\propto\dot\gamma^{1/\epsilon}$ to obtain the inverse logarithmic slope $\epsilon$ \citep{BJ09}.  Using this construction, $\epsilon=1$ corresponds to a Newtonian scaling, and smaller values of $\epsilon$ correspond to  shear thickening.  Figure \ref{fig:epsilon_phirel} shows the behavior of $\epsilon$ for several suspensions.  This is plotted vs. a normalized packing fraction $\phi/\phi_c$, where $\phi_c$ is the packing fraction obtained independently as the onset of a yield stress due to jamming.  This plot  includes different particle shapes, each with different $\phi_c$.   As $\phi$ approaches $\phi_c$ from below, $\epsilon$  approaches zero.  This implies that the viscosity curves approach the limit of a discontinuous increase in $\tau(\dot\gamma)$ as $\phi$ reaches the jamming transition.  Thus, the term `discontinuous' is only strictly descriptive in the limit of $\phi\rightarrow \phi_c$.  Most of the values of $\epsilon$ vs. $\phi/\phi_c$ in Fig.~\ref{fig:epsilon_phirel} collapse for $\phi/\phi_c \stackrel{>}{_\sim} 0.8$ (the gray shaded band), suggesting a universal scaling in this regime.  To the extent that this analogy holds,  this would be similar to a second order phase transition in which proximity to the critical point controls the strength of shear thickening, where value of the critical point is the same as the jamming transition.

The one particle type clearly behaving differently are S-shaped hooks.  This suggests that while all of the convex particle shapes collapse onto the same universal scaling, more extreme non-convex particle shapes may introduce additional effects, which are not yet understood.

 \section{Microstructure and intrinsic rheology}
\label{sec:microstructure}
 
 \subsection{Microstructure}
 
Some of the main approaches to understanding DST based on the standard paradigm in rheology, which is to relate microstructural changes to intrinsic bulk viscosities. Dilute complex fluids can be treated as a perturbation on Newtonian fluids, and early work successfully attributed shear thinning \citep{CK86} and continuous shear thickening \citep{BB85} to changes in the local microstructural arrangements of particles in terms of a structure function.  In both of these cases, the rearrangement of particles due to a change in shear rate leads to slight changes in viscous drag forces between neighboring particles.  However, the resulting changes in effective viscosity are less than a factor of two.  This is because the fundamental interactions between particles that are responsible for the measured forces do not change, and only the values of the forces change slightly due to the changing distribution of neighboring particle distances and orientations. 

A variety of observations suggest such microstructural changes do not directly produce the dramatic changes in stress, for both strong shear thinning and shear thickening systems.  For example, microstructural changes at the onset of shear thickening are not consistent.  Different microstructural changes can be observed depending on which forces are dominant for a  particular suspension; sometimes this is an order-disorder transition \citep{Ho72},  but this transition does not always occur \citep{MW02}.    Second, different stress responses can be found for similar microstructures.  For example, random particle arrangements can be found for both strong shear thinning due to entropic forces \citep{CMIC11, XRD12} or weaker shear thinning due to changes in viscous forces as particle structures rearrange \citep{CK86}.  Therefore, large changes in the viscosity do not correspond 1-to-1 to changes in microstructure.

These observations may be leading to a paradigm change in the interpretation of DST and other rheology involving large changes in viscosity, in which microstructural rearrangements may be thought of as more of a byproduct rather than the cause of changes in dominant stress scales in different regimes \citep{CMIC11, BJ11, XRD12}.


 %


 \subsection{Constitutive relation}
\label{sec:constitutive}

Taking the aforementioned considerations into account, a simple approximate constitutive equation for the rheology DST fluids can be written as \citep{BJ12}

\be
\tau = \eta_{\nu}(\phi)\dot\gamma  + \mu \tau_{conf}(\delta)+\tau_{min} \ .
\label{eqn:constitutivelaw2}
\ee

\noindent  Here $\tau_{min}$ represents the combined effect of forces that result in shear thinning at low shear rates, such as direct attractive or repulsive forces between particles, osmotic pressure, or gravity.   There is a confining stress $\tau_{conf}(\delta)$ where $\delta$ is a measure of dilation, and $\mu$ in an effective friction coefficient.  When the dilation is against a linear elastic boundary with a per-particle stiffness $k$ (Fig.~\ref{fig:stressmaxcollection}), then $\tau_{conf}(\delta) = \delta k/a^2$ \citep{BJ12}.  For simplicity, we assume in Eqn.~\ref{eqn:constitutivelaw2} that the flow speed is low so there are no inertial stresses, but it is straightforward to add that contribution \citep{BJ12}.  

Despite its simplicity, this constitutive relation still captures the basic rheology of DST fluids and the phase diagram illustrated in Fig.~\ref{fig:phasediagram}.  It relates strong variations in the effective viscosity to changes in the dominant stresses between particles. These are able to change rapidly with the shear rate in response to solid-solid frictional contacts and dilation which can be characterized simply by a global volumetric change \citep{BJ12}, rather than being sensitive to a local structure function or other details of the microstructure.  However, the dependence of dilation on shear rate is not explicit.  While more general constitutive relations such as the Pouliquen rheology \citep{BGP11} have been developed for constant normal force boundary conditions, such relations have not yet been applied to model DST systems.

\subsection{Intrinsic rheology vs. the significance of boundary conditions}

 It is interesting to note that the direct shear-rate dependence of the constitutive equation in Eq.~\ref{eqn:constitutivelaw2}  is inherently shear thinning. 
Local shear profile measurements confirmed that such local relationship can indeed be shear thinning for DST suspensions \citep{BJ12}.  The explanation for this apparent contradiction is that most of the shear stress is due to solid-solid frictional contacts and thus comes through the non-local confining stress term which is a response to the global dilation $\delta$.  One of the surprising consequences of this is that characterizing rheology  solely via local, shear-rate dependent constitutive laws or local viscosities in the bulk would miss the dramatic features of DST.

 From a hydrodynamic point of view, the large significance of the boundary conditions and difference between local and global results is unusual. In this traditional context it is more typical for the stresses to be dominated by bulk viscous or other interparticle stresses, while boundary conditions to play a smaller role, requiring only perturbative corrections to translate between the local and global rheology.   

A defining feature  of intrinsic behavior is that stresses, strains, and shear rates can describe bulk material properties independent of system size.  Typically, boundary conditions only contribute significant effects near the boundaries of continuum systems, so their contributions tend to decrease in relative importance when the system size gets larger.  On the other hand, in DST, the boundary transmits forces along solid contacts between neighboring particles in the system.  If the system as a whole is jammed (even if individual force chains exist transiently), these forces transmit all the way to opposite boundaries.  The magnitude of forces and density of particles do not decrease as they move further into the system, so the scale of the corresponding stress remains independent of the system size as for an intrinsic source of stress, even though the source of the stress comes from the boundary \citep{BZFMBDJ10}.  Thus stress, strain, and shear rate remain meaningful ways of characterizing forces, displacements, and velocities in a system-size-indpendent way for DST systems, as with any other continuum material.

\section{Dynamic phenomena associated with shear thickening} 
\label{sec:phenomenology}


While the majority of work on shear thickening has been focused on the behavior of  $\tau(\dot\gamma)$ under steady state flow conditions, dense suspension exhibit a range of remarkable  dynamic phenomena that emerge when the system is in a non-steady, transient state.  Without trying to be exhaustive, we here introduce some of these behaviors and discuss their relation to shear thickening.





\subsection{Stable fingers and holes in vibrated layers}

\begin{figure}                                                
\centerline{\includegraphics[width=3.in]{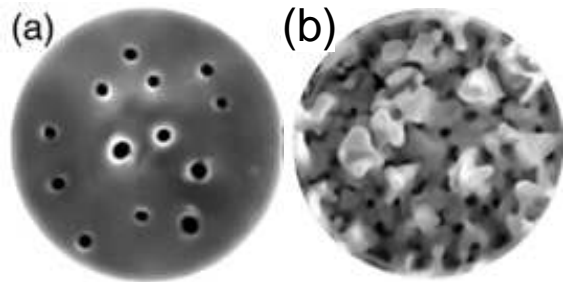}}
\caption{(a) Persistent holes and (b) dynamic fingers and holes in a vibrated layer of cornstarch in water.  Figure reproduced from \citet{MDGRS04}.}
\label{fig:vibration}                                        
\end{figure}

When vertically vibrated, a layer of Newtonian fluid can undergo instabilities that deform its free surface.  However, gravity and surface tension provide restoring forces that tend to drive transient perturbations of the fluid surface back to a flat state that minimizes the surface energy. Localized deformations of the surface that persist over many vibration cycles are therefore suppressed. Shear-thickening suspensions can exhibit behavior that  violates this rule.   Thin layers of dense suspension, flat and liquid-like at rest, under  sufficiently strong vibration develop protrusions that grow finger-like  into the third dimension. As shown by \citet{MDGRS04} these fingers and holes are among a whole family of instabilities, including also open holes that are stable under vibration but close when the driving is turned off.  More recently, it was found that for certain kinds of suspension the holes  could expand, split and replicate, similar to self-replicating spots in chemical diffusion-reaction systems \citep{XGSXJZ11, ES11}.

\citet{MDGRS04} originally attributed the holes' stability to shear thickening, but \citet{De10} later showed  that the persistence is a consequence of stress hysteresis commonly observed in dense suspensions.  Such hysteresis is is also a key aspect of equation of state models for shear thickening \citep{HAC01}; however, for structures such as holes and fingers, shear thickening per se may not be a requirement.  Indeed, recent experiments observed the very same localized structures also in vibrated emulsions that are shear thinning \citep{FBPD12}.  Simulations have also produced dynamic holes and fingers using a non-Newtonian fluid model that does not have shear thickening viscosity curve, and consistent with the conclusion of \citet{De10}, the fingers and holes only form when hysteresis is explicitly included in the model \citep{OBK13}.  These results all lead to the same conclusion: persistent, dynamic  holes and fingers are due to stress hysteresis rather than shear thickening.




\subsection{Impact resistance and solid-like behavior}

One of the best known features of shear thickening fluids is their remarkable  impact resistance.  As Fig.~\ref{fig:cornstarchpool} shows, a dense suspension can easily support the weight of a grown person running across it. To prevent an adult from sinking in, a simple estimate shows that the suspension must support on average normal stresses  in excess of about 40kPa.  Such stress levels are an order of magnitude larger than the upper limit $\tau_{max}$ for shear thickening in cornstarch and water. They are also significantly larger than most other reported $\tau_{max}$ values (Fig.~\ref{fig:stressmaxcollection}).

This situation creates a problem for explaining the observed impact resistance with any of the models discussed so far. Lubrication forces are unable to generate these high stress levels by themselves.  However, there are many recorded instances where people have been able to run across the surface of whole pools filled with a dense suspension (these are easily searchable on YouTube).  

Recent experiments have shown that the impact behavior is  linked not so much to the steady-state shear response but rather is a transient response \citep{WJ12}. Rapid normal impact onto the free surface of a dense suspension generates a  compression  of the particle sub-phase, which initiates a propagating density front that  transforms the  fluid into a temporarily jammed solid. Once this front has reached the bottom  of the system, a direct connection to  a boundary is established, which can transmit stresses back to the impacting object. In relatively shallow layers of suspension, this connection is  so solid-like that a bowling ball hitting the surface can bounce back. The presence of such recoil also indicates that there is at least some elastic energy contributed from compression of the particles in the suspension.

But even before the front reaches the bottom, or in deep systems, very large normal stresses are created simply by the fact that the jammed region is rapidly growing.  This growth was found to be proportional to the distance the impacting object pushes the suspension surface downward, producing an effect similar to the rapid growth of compacted mass in front of a shovel that is pushed into snow \citep{WRVJ13}.The front propagates downward as well as radially outward (in dense cornstarch/water suspensions about 10 times the pushing distance \citep{WJ12}) and the solid region generated in the process not only increases the inertia, but also the effective drag. As a result, dense cornstarch and water suspensions have been observed to support normal stresses up to 1 MPa regime even before the jammed region reaches a wall \citep{WJ12}.    

\begin{figure}                                                
\centerline{\includegraphics[width=3.4in]{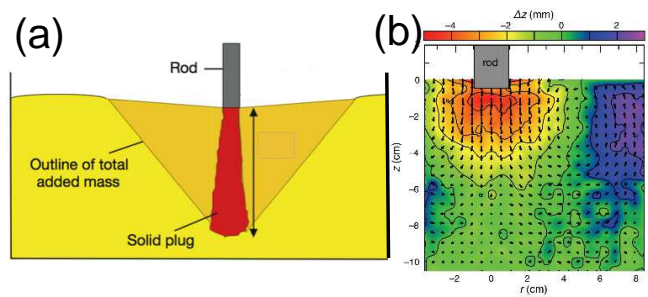}}
\caption{(a) Sketch of the solidification below a rod impacting the surface of a suspension of cornstarch and water.  The red area represents the solidified region, while the surrounding orange color represents the effective shape of the added mass.  The yellow represents the remaining liquid-phase suspension. (b) Cross-section of displacement field $\Delta z$ calculated from X-ray images of tracers inside the suspension, taken 60 ms after impact. The color scale corresponds to the size of vertical displacements. The large red/yellow region outlines the material that is forced downward by the rod.  Figure based on \citet{WJ12}.
}
\label{fig:impact}                                     
\end{figure}

Figure \ref{fig:impact} shows a sketch of the growing, solidified region, consisting of a central jammed plug surrounded by added mass that is dragged along. Direct imaging of the evolving front has so far not been achieved in a three-dimensional system, although the resulting net displacement can be reconstructed from tracking tracer particles by x-rays (Fig.~\ref{fig:impact}b).
Earlier experiments by \citet{LSZ10} had  already provided  indirect evidence for jammed, well-delineated plugs underneath large spheres that were pushed downward inside a dense suspension with a linear actuator. In particular, they  demonstrated the remarkably focused nature of the stress transmission by observing the indentations generated by the plugs on a tank bottom made from soft molding clay.   When using an elastomer (PDMS) instead of clay, similarly focused indentations were found, clearly establishing the jammed material in front of the sphere was truly solid-like; but the elastomer was also seen to relax back after a short  time, indicating the transient nature of this jammed solid.

The consequences of dynamic jamming can already be observed when an object simply sinks  into a dense suspension  \citep{KSLM11}.  In particular when the object approaches the bottom boundary, the slow down due to the growing solid-like region in front of the object can lead to a complete stop, so that the jammed solid has time to dissolve and `melt'.  The result are stop-go oscillations. The characteristic time scale for the locally hardened, jammed material to soften, seen in these settling oscillation, is of the same order of magnitude, around 25-100ms, as observed in the impact experiments  \citep{WJ12, LSZ10} and depends on a combination of suspension parameters such as packing fraction and  viscosity of the suspending fluid (the Twente group models this by Darcy flow through a porous medium \citep{KSLM11}).  This scale also may set the time delay after which a person's foot will start to sink into the suspension (Fig.\ref{fig:cornstarchpool}). On the other hand, the initiation of the jamming front and the associated  normal stress response to impact appear to be comparatively insensitive to changes in parameters such as the solvent viscosity \citep{WJ12}.

The limit of very fast impacts, with strain loading rates  up to around 100,000/s, can be probed by using a set-up in which  the sample is sandwiched between two metal cylinders instrumented with strain gauges (split Hopkinson bar), one of which is then struck by a gas-driven high-velocity anvil (striker bar) \citep{LLWG10b}.   Brownian motion is unlikely to play much of a role at these extreme loading rates. Thus, while the results of split Hopkinson bar experiments have typically been interpreted in terms of particle cluster formation, in the absence of Brownian motion this is  essentially a granular scenario and likely involves  jamming as a consequence of compression of the particle sub-phase.  Indeed, the observed stress levels of up to tens of MPa \citep{LLWG10b, JGXJYLL13} are far beyond the range of lubrication forces.  Note that stresses reported from these experiments are measured after the initial impulse from the anvil has been reflected several times back and forth between the two surfaces confining the sample (i.e., after the so-called ringing-up period).  These stresses therefore correspond to the situation after a jammed solid bridging the full sample depth has already been established.  As the stress levels increase, a transition to a regime dominated by the modulus of the particles is observed, followed by non-reversible fracturing \citep{LLWG10}.  

Even higher, ballistic impact speeds in excess of 1000m/s can be achieved with explosively launched flyer plates \citep{PH13, POLMFH13}. From experiments of this type, using SiC particles, transient shear stress levels of 0.5GPa have been inferred. This clearly indicates stress transmission via solid-solid contacts provided by the granular network of force chains and is eventually limited by the stiffness of the individual particles.  For the use of DST suspensions in protective vests or clothing the implications are that improvements are more  likely to come from optimizing the frictional nature of the particle contacts, than from tuning hydrodynamic interactions mediated by the suspending fluid \citep{KMMW09,POLMFH13}.

\begin{figure}                                                
\centerline{\includegraphics[width=3.4in]{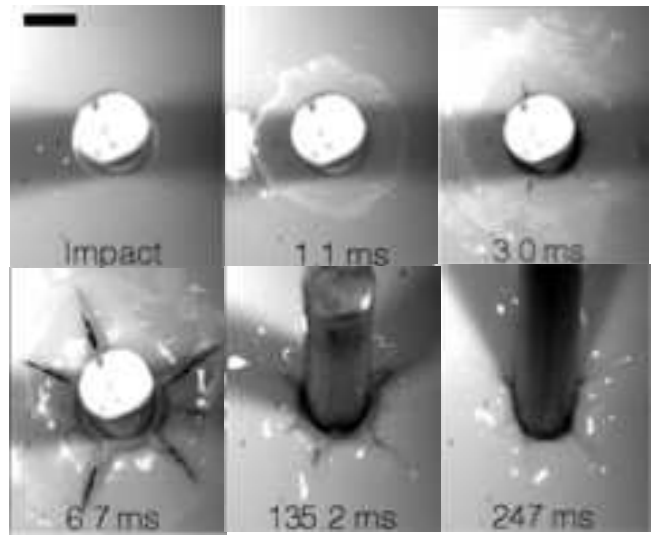}}
\caption{Time sequence of cracking of a suspension of cornstarch and water impacted by a metal rod.  Immediately after impact, a region where particles penetrate the surface appears around the impactor, followed by radial cracks within this region after a few ms. After about 100 ms, the cracks have already started to heal as the suspension becomes liquid-like again.  Scale bar: 1 cm.  The apparent change in the rod length is a visual artifact and due to tilting once the suspension starts to `melt' after impact.  Figure reproduced from \citet{RMJKS13}.
}
\label{fig:cracks}                                        
\end{figure}

Recently \citet{RMJKS13} investigated fracturing of dense suspensions at lower impact speeds by direct video imaging. In these experiments, performed with cornstarch and water at relatively low-packing-fraction ($\sim40$\%), the vertical impact of a cylindrical rod initially created the same rapidly growing dynamic jamming front observed by \citep{WJ12}, seen at the suspension surface as a radially spreading change from smooth and glossy to rough and matt. This was followed by penetration of the rod into the suspension and the appearance of cracks moving radially outward from the impact site, similar to a mode-1 fracture (Fig.~\ref{fig:cracks}).   Brittle fracture and cracking have  also been observed in shear and extensional flow of shear thickening suspensions \citep{WCR10, SBCB10}, see for example Fig. 6.


While it has long been claimed that dense cornstarch suspensions have elastic properties, few clear quantitative observations have been reported.   In tensile measurements long filaments have been observed that appear fluid-like but exhibit viscoelastic recoil with forces provided by surface tension, not in response to dilation but by the retraction of the interstitial liquid \citep{SBCB10}.  \citet{WJ12} reported bouncing of objects impacting the surface of a cornstarch suspension.   Steady shear experiments with a small oscillatory component on top of a constant shear stress demonstrated the existence of an elastic modulus in the shear thickening regime \citep{Ru13}.  The scenario emerging from all of these observations is that shear thickening suspensions exhibit a number of solid-like properties, and hints of elastic properties, although it remains to be determined what determines their stiffness and strength.


 \section{Conclusions}
 
 There have been great strides in understanding Discontinous Shear Thickening in the past two decades since Barnes' review.  We can now make several well-supported conclusions about the mechanisms for DST:  

\begin{enumerate}

\item The stress at the onset of shear thickening $\tau_{min}$ can be described well by very simple force balance models.  $\tau_{min}$ is on the scale of the largest source of stress that opposes shear of particles past each other -- this largest stress depends on which forces are dominant for given particle size and other material properties of the suspensions.  This conclusion is independent of the particular structural transition that occurs at the onset of shear thickening.  In different shear thickening suspensions, there is evidence for the  transition being hydroclustering \citep{CMIC11}, an order-disorder transition \citep{Ho74}, and dilatancy \citep{BJ12}, but all of these microstructural changes can be associated with the same onset stress scalings.
 
\item Frustrated dilatancy against a stiff boundary is required at least for the largest increases in stresses observed in shear thickening suspensions.  Lubrication forces are not strong enough to provide the largest stresses observed in suspensions of cornstarch in water among others, nor can they explain the strong coupling between shear and normal stresses and dependence on the boundary conditions.   

\item The upper bound of the shear thickening regime, $\tau_{max}$, is limited by the weakest confining stress in response to dilation in the system and provides a convenient signature of the source of confining stress for DST.  In most experiments, this is usually surface tension at the fluid-air interface, which can be magnified by orders-of-magnitude due to particles poking through the surface in response to dilation \citep{BJ12}.  In other cases, the limiting confining stress may be particle stiffness, or the stiffness of a solid wall at the boundary \citep{BJ12, WB09, OH10}.  

 \item One of the most puzzling questions that had been left open from Barnes' review was why don't  all dense suspensions exhibit shear thickening, since the proposed mechanisms should be generally applicable to all suspensions.  The solution is in the stress scales.  If the weakest confining stress in response to dilation ($\tau_{max}$) is weaker than or comparable to the strongest stress opposing shear between neighboring particles ($\tau_{min}$) , then there is no significant potential increase in stress with increasing shear rate to produce shear thickening.  Some examples that do not shear thicken are suspensions with strong attractions between particles that behave as yield stress fluids.  Other examples include very soft particles as in emulsions or foams, where particles can deform easily to shear past each other without pressing hard against the boundaries.
 
 \item The constitutive relation for $\tau(\dot\gamma)$ is not intrinsic, bulk behavior characterized by the local microstructure as is the standard expectation for complex fluids.  Instead, $\tau(\dot\gamma)$ is determined by the boundary conditions in response to the global structural change of dilation.
 
 \end{enumerate}
 
Recent years have also brought a re-evaluation of many of the dynamical phenomena that have long been associated with shear thickening.   This has led to several surprises. For example, the formation of persistent fingers and holes in vibrated layers of shear thickening fluid cannot explained by a shear thickening $\tau(\dot\gamma)$. In fact,  these persistent structures have now been observed also in non-shear thickening fluids \citep{FBPD12}, and are better explained as a consequence of hysteresis in the rheology \citep{De10, OBK13}.  The strong impact resistance of shear thickening fluids is better explained as a result of transiently jammed solid regions forming in front of the impact \citep{WJ12}, which is related to shear thickening but not characterized by the same $\tau(\dot\gamma)$ relation.

 
{\bf Open Questions and opportunities:}  The rich set of behaviors discussed in this article exists in a region of parameter space that is outside the regime traditionally investigated by either the complex fluids or the granular materials communities. Dense suspensions, as has become clear, cannot be thought of as simple extensions of the dilute limit of a few particles in a liquid. Conversely, adding interstitial liquid to a dry granular material introduces qualitatively new effects, not the least of which is the confining role of surface tension.  In the parameter region relevant to DST, which has been the focus of our review, the suspension consists roughly 50/50 of particles and of liquid, and a full treatment has to consider both.  It should therefore not come as a surprise that, despite all the advances, there remain many outstanding  problems and opportunities for further research.

In particular,  while we can identify general scaling arguments for the onset of shear thickening, there is not yet agreement on a general model that explains the onset of DST with direct supporting evidence.  It remains to be seen if there is a way to reconcile the different microstructural mechanisms into a single general model, or if the field will remain in support of several distinct models applicable to different types of suspension.   A further challenge for such models is to quantitatively predict the steep slope of viscosity curve and its evolution with packing fraction.  This probably requires a constitutive relation that relates dilation, normal stress, and shear rate in DST systems.

Shear thickening is a phenomenon which shares many properties of jammed systems. We labeled this larger set of behaviors ``dynamic jamming", mindful not only of the connections to but also of differences with a more static jammed state that has not yielded to shear.  An extended formalism for jamming that would apply to the variety of dynamic systems and transient behaviors would be an important contribution to the field of jamming.  The fact that suspensions can be prepared across a wide range of packing fractions and the ability to tune the particle interactions offers further possibilities for studying jamming.  
 
 It has not yet been demonstrated how the dynamic impact response of shear thickening systems can be explained in a detailed, quantitative manner that captures the particle microstructure as well all of the observed phenomena, such as  rolling and bouncing of objects and, of course, running on the surface. A general explanation for the elastic- and solid-like behaviors, and how they might depend on the details of the particle properties and boundary conditions remains to be worked out.
 
All of these challenges and opportunities relate to the fundamental science associated with the behavior of dense suspensions.  However, better understanding is critical also for enhancing flow and preventing clogging during the industrial processing of dense suspensions, and it has already begun to enable the design of new applications, specifically materials for improved impact dissipation.


 \section{Acknowledgements}
 
We thank Sid Nagel, Martin van Hecke, Wendy Zhang for thoughtful discussions.  We thank Benjamin Allen for the photograph in Fig.~\ref{fig:cornstarchpool}.  This work was supported in part by the National Science Foundation CAREER program under grant no. CBET-1255541 (E.B.) and by the US Army Research Office through grant no. W911NF-12-1-0182 (H.M.J.).  



\begin{thebibliography}{}



\bibitem[Aradian and Cates(2006)]{AC06} Aradian, A., and M.E. Cates, ``Minimal model for chaotic shear banding in shear thickening fluids," Phys. Rev. E {\bf 73}, 041508 (2006).

\bibitem[Bagnold(1954)]{Ba54} Bagnold, R.A., ``Experiments on a gravity-free dispersion of large solid spheres in a newtonian fluid under shear," Proc. Royal Soc. London A: Math. and Phys. Sci. {\bf 225} (1160), 49  (1954).

\bibitem[Barentin and Liu(2001)]{BL01}Barentin, C. and A. J. Liu, Europhys. Lett. ``Shear thickening in dilute solutions of wormlike micelles" {\bf 55} (3), 432 (2001).

\bibitem[Barnes(1989)]{Ba89} Barnes, H.A.,``Shear-thickening (``Dilatancy") in suspensions of nonaggregating solid particles dispersed in Newtonian liquids," J. Rheology {\bf 33} (2), 329 (1989).

\bibitem[Bender and Wagner(1996)]{BW96} Bender, J. and  N. Wagner, ``Reversible shear thickening in monodisperse and bidisperse colloidal dispersions," J. Rheology {\bf 40} (5), 899 (1996).

\bibitem[Bergenholtz et al.(2002)]{BBV02} Bergenholtz, J., J.F. Brady, and M. Vicic, ``The non-Newtonian rheology of dilute colloidal suspensions," J. Fluid Mech. {\bf 456}, 239 (2002).

\bibitem[Bertrand et al.(2002)]{BBS02} Bertrand, E., J. Bibette, and V{\'e}ronique Schmitt, ``From shear thickening to shear-induced jamming," Phys. Rev. E {\bf 66}, 060401(R) (2002).

\bibitem[Bi et al.(2011)]{BZCB11} Bi, D., J. Zhang B. Chakraborty, and R. P. Behringer, ``Jamming by shear," Nature {\bf 480} (7377) 355-358 (2011).

\bibitem[White et al.(2010)]{WCR10}White, E.E.B., M. Chellamuthu, and J.P. Rothstein, ``Extensional rheology of a shear-thickening cornstarch and water suspension," Rheol. Acta. {\bf 49} 119 (2010).

\bibitem[Bocquet et al.(2001)]{BLSLG01} Bocquet, L., W. Losert, D. Schalk, T.C. Lubensky, and J.P. Gollub, ``Granular shear flow dynamics and forces: experiment and continuum theory," Phys. Rev. E {\bf 65}, 011307 (2001).

\bibitem[Boersma et al.(1991)]{BBLS91} Boersma, W.H., P.J. M. Baets, J. Laven, and H.N. Stein, ``Time-dependent behavior and wall slip in concentrated shear thickening dispersions," J. Rheol. {\bf 35} (6), 1093 (1991).

\bibitem[Boersma et al.(1990)]{BLS90} Boersma, W.H., J. Laven, H.N. Stein, ``Shear thickening (dilatancy) in concentrated dispersions," AIChE {\bf 36}, 321 (1990).

\bibitem[Boyer et al.(2011)]{BGP11} Boyer, F., E. Guazzelli, and O. Pouliquen, ``A constitutive law for dense granular flows,"  Phys. Rev. Lett. {\bf 107}, 188301 (2011).

\bibitem[Brady and Bossis(1985)]{BB85} Brady, J.F. and G. Bossis, ``The rheology of concentrated suspensions of spheres in simple shear flow by numerical simulation," J. Fluid Mech. {\bf 155}, 105 (1985).

\bibitem[Brady and Vicic(1995)]{BV95} Brady, J.F. and M. Vicic, ``Normal stresses in colloidal dispersions," J. Rheol. {\bf 39} (3), 545 (1995).

\bibitem[Brown and Jaeger(2009)]{BJ09} Brown, E and H.M. Jaeger, ``Dynamic jamming point for shear thickening suspensions," Phys. Rev. Lett {\bf 103} 086001 (2009).

\bibitem[Brown et al.(2010a)]{BFOZMBDJ10} Brown, E., N. A. Forman, C. S. Orellana, H. Zhang, B. W. Maynor, D. E. Betts, J. M. DeSimone, and H. M. Jaeger, ``Generality of shear thickening in suspensions," Nature: Materials {\bf 9} (3) 220-224 (2010a).

\bibitem[Brown et al.(2010b)]{BZFMBDJ10}  Brown, E. H. Zhang, N. A. Forman, B. W. Maynor, D. E. Betts, J. M. DeSimone, and H. M. Jaeger,  ``Shear thickening in densely packed suspensions of spheres and rods confined to few layers," J. Rheol. {\bf 54} (5) 1023 (2010b).

\bibitem[Brown et al.(2011)]{BZFMBDJ11}Brown, E. H. Zhang, N. A. Forman, B. W. Maynor, D. E. Betts, J. M. DeSimone, and H. M. Jaeger,  ``Shear thickening and jamming in densely packed suspensions of different particle shapes,"  Phys.~Rev.~E. {\bf 84} 031408 (2011).

\bibitem[Brown and Jaeger(2011)]{BJ11} Brown, E. and H. M. Jaeger. ``Through Thick and Thin," Science {\bf 333}, 1230 (2011).

\bibitem[Brown and Jaeger(2012)]{BJ12} Brown, E. and H. M. Jaeger. ``The role of dilation and confining stress in shear thickening of dense suspensions." J. Rheology 56 (4), 875-923 (2012).

\bibitem[Cates et al.(2005b)]{CAS05} Cates, M.E., R. Adhikari, and K. Stratford, ``Colloidal arrest by capillary forces," J. Phys.: Condens. Matter {\bf 17}, S2771 (2005).

\bibitem[Cates et al.(2005a)]{CHH05} Cates, M.E., M.D. Haw and C.B. Holmes, ``Dilatancy, jamming, and the physics of granulation," J. Phys: Condens. Matter {\bf 17} S2517 (2005).

\bibitem[Cates et al.(1998)]{CWBC98} Cates, M.E., J.P. Wittmer, J.-P. Bouchaud, and P. Claudin, ``Jamming, force chains, and fragile matter," Phys. Rev. Lett. {\bf 81} (9), 1841 (1998). 

\bibitem[Chellamuthu et al.(2009)]{CAR09} Chellamuthu, M., E.M. Arndth, and J.P. Rothstein, ``Extensional rheology of shear thickening nanoparticle suspensions," Soft Matter {\bf 5} (10), 2117 (2009).

\bibitem[Cheng et al.(2011)]{CMIC11} Cheng, X., J.H. McCoy, J.N. Israelachvili, and I. Cohen, ``Imaging the microscopic structure of shear thinning and thickening colloidal suspensions,"  Science {\bf 333} 1276-1279 (2011).

\bibitem[Choi and Krieger(1986)]{CK86} G.N. Choi, I.M. Krieger, ``Rheological studies on sterically stabilized model dispersions of uniform colloidal spheres: II. Steady-shear viscosity," J. Colloid and Interface Science {\bf 113} (1), 101 (1986).

\bibitem[Corwin et al.(2005)]{CJN05} Corwin, E.I., H.M. Jaeger and S.R. Nagel, ``Structural signature of jamming in granular media," Nature {\bf 435}, 1075 (2005).

\bibitem[Deegan(2010)]{De10} Deegan, R.D., ``Stress hysteresis as the cause of persistent holes in particulate suspensions," Phys. Rev. E {\bf 81}, 036319 (2010).

\bibitem[Donnelly and Simon(1960)]{DS60} Donnelly, R.J. and N.J. Simon, ``An empirical torque relation for supercritical flow between rotating cylinders," J. Fluid. Mech. {\bf 7}, 401 (1960).


\bibitem[Deboeuf et al(2008)]{DGMYM09}Deboeuf, A., G. Gauthier, J. Martin, Y. Yurkovetsky, and J. F. Morris, ``Particle pressure in a sheared suspension: a bridge from osmosis to granular dilatancy," Phys.~Rev.~Lett. {\bf 102} 108301 (2009).

\bibitem[Ebata and Sano(2011)]{ES11}Ebata, H. and M. Sano, ``Self-Replicating Holes in a Vertically Vibrated Dense Suspension," Physical Review Letters {\bf 107}, 088301 (2011).

\bibitem[Egres and Wagner(2005)]{EW05} Egres, R.G. and N. J. Wagner, ``The rheology and microstructure of acicular precipated calcium carbonate colloidal suspensions through the shear thickening transition," J. Rheol. {\bf 49} (3), 719 (2005).

\bibitem[Egres et al.(2006)]{ENW06} Egres, R.G., F. Nettesheim, and N.J. Wagner, ``Rheo-SANS investigation of acicular-precipitated calcium carbonate colloidal suspensions through the shear thickening transition," J. Rheol. {\bf 50} (5), 685 (2006).

\bibitem[Falcon et al.(2012)]{FBPD12} Falcon, C., J. Bruggeman, M. Pasquali, and R. D. Deegan, ``Localized structures in vibrated emulsions," Europhysics Letters {\bf 98} (2012).

\bibitem[Fall et al.(2009)]{FBOB09} Fall, A., F. Bertrand, G. Ovarlez, and D. Bonn, ``Yield stress and shear banding in granular suspensions," Phys. Rev. Lett. {\bf 103}, 178301 (2009).

\bibitem[Fall et al.(2008)]{FHBOB08} Fall, A., N. Huang, F. Bertrand, G. Ovarlez, and D.Bonn, ``Shear thickening of cornstarch suspensions as a reentrant jamming transition," Phys. Rev. Lett. {\bf 100}, 018301 (2008).

\bibitem[Fall et al.(2010)]{FLBBO10} Fall, A., A. Lema{\^i}tre, F. Bertrand, D. Bonn, G. Ovarlez, ``Continuous and discontinuous shear thickening in granular suspensions," submitted to Phys. Rev. Lett. (2010).

\bibitem[Farr et al.(1997)]{FMB97} Farr, R.S., J.R. Melrose, and R.C. Ball, ``Kinetic theory of jamming in hard-sphere startup flows," Phys. Rev. E {\bf 55} (6), 7203 (1997).

\bibitem[Frankel and Acrivos(1967)]{FA67}Frankel, N.A., and A. Acrivos, ``On the viscosity of a concentrated suspension of solid spheres," Chem. Eng. Sci. {\bf 22} 847-853 (1967).

\bibitem[Frith et al.(1996)]{FHBM96} Frith, W.J., P. d'Haene, R. Buscall, J. Mewis,  ``Shear thickening in model suspensions of sterically stabilized particles," J. Rheology {\bf 40} (4), 531 (1996).

\bibitem[Freundlich \& Roder(1938)]{FR38}Freundlich, H., and Roder, H. C., Trans. Faraday Sot., {\bf 34}, 308 (1938).

\bibitem[Galley and Puddington(1943)]{GP43} Gallay, W. and I.E. Puddington, Can. J. Res. C {\bf 21}, 179 (1943).

\bibitem[Gardiner et al.(1998)]{GDJC98}  B.S. Gardiner, B.Z. Dlugogorski, G.J. Jameson, R.P. Chhabra, J. Rheol {\bf 42} (6), 1437 (1998).

\bibitem[Goldman et al.(1967)]{GCB67} Goldman, A.J., R.G. Cox, and H. Brenner, ``Slow viscous motion of a sphere parallel to a plane wall -- Couette flow," Chem. Eng. Sci. {\bf 22}, 653 (1967).

\bibitem[Gopalikrishnan and Zukoski(2004)]{GZ04} Gopalakrishnan, V., and C.F. Zukoski, ``Effect of attractions on shear thickening in dense suspensions," J. Rheol. {\bf 48} (6), 1321 (2004).

\bibitem[Grebenkov(2008)]{GCNC08} Grebenkov, D.S., M.P. Ciamarra, M. Nicodemi, and A. Coniglio, ``Flow, ordering, and jamming of sheared granular suspensions," Phys. Rev. Lett. {\bf 100}, 078001 (2008).

\bibitem[Head et al.(2001)]{HAC01} D.A. Head, A. Adjari, and M.E. Cates, ``Jamming, hysteresis, and oscillation in scalar models for shear thickening," Phys. Rev. E {\bf 64}, 061509 (2001).

\bibitem[H{\'e}braud and Lootens(2005)]{HL05} H{\'e}braud, P. and D. Lootens,  ``Concentrated suspensions under flow: shear-thickening and jamming," Mod. Phys. Lett. B {\bf 19} (13-14), 613 (2005). 

\bibitem[Hoffman(1972)]{Ho72} Hoffman, R. L., ``Discontinuous and dilatant viscosity behavior in concentrated suspensions: observation of a flow instability," Trans. Soc. Rheol. {\bf 16}, 155 (1972).

\bibitem[Hoffman(1974)]{Ho74} Hoffman, R.L., ``Discontinuous and dilatant viscosity behaivior in concentrated suspensions II. Theory and experimental tests," J. Colloid Interface Sci. {\bf 46} 491 (1974).

\bibitem[Hoffman(1982)]{Ho82} Hoffman, R.L., ``Discontinuous and dilatant viscosity behavior in concentrated suspensions III: necessary conditions for their occurrence in viscometric flows,"  Advances in Colloid and Interface Sci. {\bf 17} 161 (1982).

\bibitem[Hoffman(1998)]{Ho98} Hoffman, R.L., ``Explanations for the cause of shear thickening in concentrated colloidal suspensions," J. Rheol. {\bf 42} (1), 111 (1998).

\bibitem[Hofmann et al.(1991)]{HRH91} Hofmann S., Rauscher, A. and Hoffmann H., ``Shear Induced Micellar Structures," Ber. Bunsenges. Phys. Chem., {\bf 95} (1991) 153

\bibitem[Holmes et al.(2003)]{HFC03} Holmes, C.B., M. Fuchs, and M.E. Cates, ``Jamming transitions in a schematic model of suspension rheology," Europhys. Lett. {\bf 63} (2), 240 (2003).

\bibitem[Holmes et al.(2005)]{HCFS05} Holmes, C.B., M. E. Cates, M. Fuchs, and P. Sollich, ``Glass transitions and shear thickening suspension rheology," J. Rheol. {\bf 49} (1), 237 (2005).

\bibitem[Hu et al.(1998)]{HBMP98} Hu, Y.T., P. Boltenhagena, E. Matthys, and D. J. Pine, ``Shear thickening in low-concentration solutions of wormlike micelles. II. Slip, fracture, and stability of the shear-induced phase," J. Rheology {\bf 42} (5), 1209 (1998).

\bibitem[Hunt et al.(2002)]{HZCB02}Hunt, M.L., R. Zenit, C.S. Campbell, and C.E. Brennan, ``Revisiting the 1954 suspension experiments of R. A. Bagnold," J. Fluid Mech. {\bf 452} 1-24 (2002).

\bibitem[Jaeger et al.(1996)]{JNB96} Jaeger, H.M., S.R. Nagel, R.P.Behringer, ``Granular, solids, liquids, and gases," Rev. Mod. Phys. {\bf 68}, 1259 (1996).

\bibitem[Janssen(1895)]{Janssen1895} Janssen, Z., ``Experiments on corn pressure in silo cells," Verein Deutsch. Ing. {\bf 39}, 1045 (1895).

\bibitem[Jerkins et al.(2008)]{JSSSSA08} Jerkins, M., M. Schr{\"o}ter, H.L. Swinney, T.J. Senden, M. Saadatfar, T. Aste, ``Onset of mechanical stability in random packings of frictional spheres," Phys. Rev. Lett. {\bf 101}, 018301 (2008).

\bibitem[Jiang et al.(2013)]{JGXJYLL13}Jiang, W.F., X. L. Gong, S. H. Xuan, W. Q. Jiang, F. Ye, X. F. Li, and T. X. Liu, ``Stress pulse attenuation in shear thickening fluid," Applied Physics Letters {\bf 102}, 101901 (2013).

\bibitem[Johnson et al.(2013)]{JHMK13}Johnson, B.L., M. R. Holland, J.G. Miller, and J.I. Katz, ``Ultrasonic attenuation and speed of sound of cornstarch suspensions," J. Acoustical Society of America {\bf 133} (3) 1399-1403 (2013).

\bibitem[Jolly and Bender(2006)]{Lord} Jolly, M.R. and J.W. Bender, US patent application No. 20060231357.

\bibitem[Jomha and Reynolds(1993)]{JR93} Jomha, A.I. and P.A. Reynolds, ``An experimental study of the first normal stress difference -- shear stress relationship in simple shear flow for concentrated shear thickening suspensions," Rheol. Acta {\bf 32} 457 (1993).

\bibitem[Kabla and Senden(2009)]{KS09} Kabla, A.J. and T.J. Senden, ``Dilatancy in slow granular flows," Phys. Rev. Lett. {\bf 102}, 228301 (2009).

\bibitem[Kalman et al.(2009)]{KMMW09} Kalman, D.P., R.L. Merrill, and N.J. Wagner, ``Effect of Particle Hardness on the Penetration Behavior of Fabrics Intercalated with Dry Particles and Concentrated Particle-Fluid Suspensions," Applied Materials \& Interfaces {\bf 1} (11), 2602 (2009).


\bibitem[Koos and Willenbacher(2011)]{KW11} Koos, E. and N. Willenbacher, ``Capillary forces in suspension rheology," Science {\bf 331} (6019), 897 (2011).


\bibitem[Lambe and Whitman(1969)]{LW69} Lambe, T.W. and R.V. Whitman, (1969) Soil Mechanics (John Wiley and Sons, New York).

\bibitem[Larsen et al.(2010)]{LKZW10} Larsen, R.J., J.-W. Kim, C. F. Zukoski, and D.A. Weitz, ``Elasticity of dilatant particle suspensions during flow," Phys. Rev. E {\bf 81}, 011502 (2010).

\bibitem[Laun(1984)]{Laun84} Laun, H.M., ``Rheological properties of aqueous polymer dispersions," Die Angewandte Makromolekulare Chemie {\bf 123}, 335 (1984).

\bibitem[Laun(1994)]{Laun94} Laun, H.M., ``Normal stresses in extremely shear thickening polymer dispersions," J. Non-Newtonian Fluid Mech {\bf 54} 87 (1994).

\bibitem[Laun et al.(1991)]{LBS91} Laun, H.M., R. Bung, and F. Schmidt, ``Rheology of extremely shear thickening polymer dispersions (passively viscosity switching fluids)," J. Rheol. {\bf 35} (6), 999 (1991).

\bibitem[Lee and Wagner(2006)]{LW06} Lee, Y.S. and N.J. Wagner, ``Rheological properties and small-angle neutron scatting of a shear thickening, nanoparticle dispersion at high shear rates," Ind. Eng. Chem. Res. {\bf 45}, 7015 (2006).

\bibitem[Lee et al.(2003)]{LWW03} Lee, Y.S., E.D. Wetzel, and N.J Wagner, ``The ballistic impact characteristics of Kevlar-woven fabrics impregnated with a colloidal shear thickening fluid," J. Materials Sci. {\bf 38}, 2825 (2003).

\bibitem[Lenoble et al.(2005)]{LSP05}Lenoble, M., P. Snabre, and B. Pouligny, ``The flow of a very concentrated slurry in a parellel-plate device: influence of gravity," Phys. Fluids {\bf 17}, 073303 (2005).

\bibitem[Lim et al.(2010a)]{LLWG10}Lim, A.S., S.L. Lopatnikov, N.J. Wagner, and J.W. Gillespie, ``An experimental investigation into the kinematics of a concentrated hard-sphere colloidal suspension during Hopkinson bar evaluation at high stresses," J. Non-Newtonian Fluid Mech. {\bf 165} (19), 1342 (2010).

\bibitem[Lim et al.(2010b)]{LLWG10b}Lim, A.S., S.L. Lopatnikov, N.J. Wagner, and J.W. Gillespie, ``Investigating the transient response of a shear thickening fluid using the split Hopkinson pressure bar technique," Rheologica Acta {\bf 49}, 879-890 (2010).

\bibitem[Liu and Pine(1996)]{LP96}	Liu C.-H. and Pine D. J., ``Shear-Induced Gelation and Fracture in Micellar Solutions," Phys. Rev. Lett., 77 (1996) 2121.

\bibitem[Liu and Nagel(1998)]{LN98} Liu, A.J. and S. R. Nagel, ``Jamming is not just cool any more," Nature {\bf 396}, 21 (1998).

\bibitem[Liu and Nagel(2010)]{LN10} Liu, A.J. and S. R. Nagel, ``The Jamming Transition and the Marginally Jammed Solid," Annual Review of Condensed Matter Physics {\bf1}, 347-369 (2010).

\bibitem[Liu et al.(2010)]{LSZ10} B. Liu, M. Shelley, and J. Zhang, ``Focused Force Transmission through an Aqueous Suspension of Granules,"  Phys. Rev. Lett.Ê105,Ê188301Ê(2010).

\bibitem[Loimer et al.(2002)]{LNS02} Loimer, T., A. Nir, and R. Semiat, `` Shear-induced corrugation of free interfaces in concentrated suspensions," J. Non-Newtonian Fluid Mech. {\bf 102}, 115 (2002).

\bibitem[Lootens et al.(2003)]{LDH03}  Lootens, D., H. Van Damme, and P. H{\'e}braud, ``Giant stress fluctuations at the jamming transition," Phys. Rev. Lett. {\bf 90}, No. 17, 178301 (2003). 

\bibitem[Lootens et al.(2005)]{LDHH05} Lootens, D., H. Van Damme, Y. H{\'e}mar, and P. H{\'e}braud,  ``Dilatant flow of concentrated suspensions of rough particles," Phys. Rev. Lett. {\bf 95}, 268302 (2005).

\bibitem[Majmudar and Behringer(2005)]{MB05} Majmudar, T.S. and R.P. Behringer, ``Contact force measurements and stress-induced anisotropy in granular materials," Nature {\bf 435}, 1079 (2005).

\bibitem[Maranzano and Wagner(2001a)]{MW01a} Maranzano, B.J., and N. J. Wagner, ``The effects of particle size on reversible shear thickening of concentrated colloidal suspensions," J. Chem. Phys. {\bf 114} (23),  10514 (2001a).

\bibitem[Maranzano and Wagner(2001b)]{MW01b} Maranzano, B.J. and N. J. Wagner, ``The effects of interparticle interactions and particle size on reversible shear thickening: hard-sphere colloidal dispersions," J. Rheol. {\bf 45} (5), 1205 (2001b).

\bibitem[Maranzano and Wagner(2002)]{MW02}  Maranzano, B.J. and N. J. Wagner, ``Flow-small angle neutron scattering measurements of colloidal dispersion microstructure evolution through the shear thickening transition," J. Chem. Phys. {\bf 117} (22),  10291 (2002).

\bibitem[Melrose and Ball(2001a)]{MB04a}  Melrose, J.R. and R.C. Ball,  ``Continuous shear thickening transitions in model concentrated colloids -- the role of interparticle forces," J. Rheol. {\bf 48} (5), 937 (2004a).

\bibitem[Melrose and Ball(2004b)]{MB04b} Melrose, J.R., R. C. Ball, ``Contact networks in continuously shear thickening colloids," J. Rheology {\bf 48} (5), 961 (2004b).


\bibitem[Merkt et al.(2004)]{MDGRS04} Merkt, F.S., R.D. Deegan, D.I. Goldman, E.C. Rericha, and H.L. Swinney, ``Persistent Holes in a Fluid," Phys. Rev. Lett. {\bf 92} (18), 184501 (2004).

\bibitem[Metzner and Whitlock(1958)]{MW58} Metzner, A.B., and M. Whitlock, ``Flow behavior of concentrated (Dilatant) suspensions," Trans. Soc. Rheol. {\bf 11}, 239 (1958).

\bibitem[Miskin and Jaeger(2012)]{MJ12}Miskin,M. Z., and H. M. Jaeger, ``Droplet formation and scaling in dense suspensions," Proc. Nat. Acad. Sci. {\bf 109}, 4389-4394 (2012).

\bibitem[Mueth et al.(2000)]{MDKENJ00} Mueth, D.M., G.F. Debregeas, G.S. Karczmar, P.J. Eng, S.R. Nagel, and H.M. Jaeger,  ``Signatures of granular microstructure in dense shear flows," Nature {\bf 406}, 385 (2000).

\bibitem[Mueth et al.(1998)]{MJN98} Mueth, D.M., H.M. Jaeger and S.R. Nagel, ``Force distribution in a granular medium," Phys. Rev. E {\bf 57} (3), 3164 (1998).

\bibitem[Nakanashi et al.(2012)]{NNM12} Nakanishi, H, S. Nagahiro, and N. Mitarai, ``Fluid dynamics of dilatant fluid," Phys. Rev. E {\bf 85}, 011401 (2012).

\bibitem[Nazockdast and Morris(2012)]{NM12}Nazockdast, E., and J. F. Morris, ``Microstructural theory and the rheology of concentrated colloidal suspensions,' J. Fluid Mech. {\bf 713} 42-452 (2012).

\bibitem[Neuville et al.(2012)]{NBPVBM12}Neuville, M., G. Bossis, J. Persello, O. Volkova, P. Boustingory, and M. Mosquet, ``Rheology of a gypsum suspension in the presence of different superplasticizers," J. Rheol {\bf 56}(2), 435-451 (2012).

\bibitem[Nott and Brady(1994)]{NB94} Nott, P. R., and J.F. Brady, ``Pressure-driven flow of suspensions: simulation and theory," J. Fluid Mech. {\bf 275} 157 (1994).

\bibitem[Nordstrom et al.(2010)]{NVABZYGD10} Nordstrom, K.N., E. Verneuil, P.E. Arratia, A. Basu, Z.Zhang, A.G. Yodh, J.P.Gollub, and D.J. Durian, ``Microfluidic rheology of soft colloids above and below jamming," Phys. Rev. Lett. {\bf 105} 175701 (2010).

\bibitem[O'Brien and Mackay(2000)]{OM00}  O'Brien, V.T. and M.E. Mackay, ``Stress components and shear thickening of concentrated hard sphere suspensions," Langmuir {\bf 16}, 7931 (2000).

\bibitem[O'Hern et al.(2002)]{OLLN02} O'Hern, C.S., S.A. Langer, A.J. Liu, and S.R. Nagel, ``Random packings of frictionless particles," Phys. Rev. Lett. {\bf 88} (7), 075507 (2002).

\bibitem[O'Hern et al.(2003)]{OSLN03} O'Hern, C.S., L. E. Silbert, A.J. Liu, and S.R. Nagel, ``Jamming at zero temperature and zero applied stress: the epitome of disorder," Phys. Rev. E {\bf 68} (1), 011306 (2003).

\bibitem[Onoda and Liniger(1990)]{OL90} Onoda, G.Y. and E. G. Liniger, ``Random loose packings of uniform sphers and the dilatancy onset," Phys. Rev. Lett. {\bf 64} (22), 2727 (1990).

\bibitem[Osuji et al.(2008)]{OKW08} Osuji, C.O., C. Kim, and D. A. Weitz, ``Shear thickening and scaling of the elastic modulus in a fractal colloidal system with attractive interactions," Phys. Rev. E {\bf 77}, 060402 (2008).

\bibitem[Otsuki and Hayakawa(2010)]{OH10} Otsuki, M. and H. Hayakawa, ``Critical scaling near jamming transition for frictional particles," arXiv:1006.3597 (2010).

\bibitem[Ozgen et al.(2013)]{OBK13}Ozgen, O., E. Brown and M. Kallman, ``Simulating shear thickening fluids," submitted to SIGGRAPH.

\bibitem[Petel and Higgins(2010)]{PH13}Petel, O. E. and A. J. Higgins, ``Shock wave propagation in dense particle suspensions," J. Appl. Phys. {\bf 108}
 (2010).

\bibitem[Petel et al.(2013)]{POLMFH13}Petel, O. E., S. Ouellet, J. Loiseau, B. J. Marr, D. L. Frost, and A. J. Higgins, ``The effect of particle strength on the ballistic resistance of shear thickening fluids," Applied Physics Letters {\bf 102}, 064103 (2013).

\bibitem[Prasad and Kyt{\"o}maa(1995)]{PK95} Prasad, D. and H.K. Kyt{\"o}maa, ``Particle stress and viscous compaction during shear of dense suspensions," Int.~J.~Multiphase Flow {\bf 21} (5), 775 (1995).

\bibitem[Raghavan et al.(2000)]{RWK00} Raghavan, S.R., H.J. Walls, and S.A. Khan, ``Rheology of silica dispersions in organic liquids: new evidence for solvation forces dictated by hydrogen bonding," Langmuir {\bf 16}, 7920 (2000).

\bibitem[Reynolds(1885)]{Re1885}  Reynolds, O., ``On the dilatancy of media composed of rigid particles," Phil. Mag.  {\bf 20}, 469 (1885).

\bibitem[Rubio-Hernandez(2013)]{Ru13} Rubio-Hern{\'a}ndez, F.J., ``In situ analysis of shear-induced microstructures," submitted. 

\bibitem[Roch{\'e} et al.(2013)]{RMJKS13} Roch{\'e}, M., E. Myftiu, M.C. Johnston, P. Kim, and H. A. Stone, ``Dynamic Fracture of Nonglassy Suspensions," Phys. Rev. Lett. {\bf 110} 148304 (2013).

\bibitem[Schlichting(1960)]{Schlichting} Schlichting, H. Boundary Layer Layer Theory. 4th ed., McGraw-hill Book Co., 1960.

\bibitem[Shenoy and Wagner(2005)]{SW05} Shenoy, S.S. and N.J. Wagner, ``Norman J. Wagner Influence of medium viscosity and adsorbed polymer on the reversible shear thickening transition in concentrated colloidal dispersions,"  Rheol. Acta. {\bf 44}, 360 (2005).

\bibitem[Shenoy et al.(2003)]{SWB03} Shenoy, S.S., N.J. Wagner,  and J.W. Bender, ``E-FiRST: Electric field responsive shear thickening fluids," Rheol Acta {\bf 42}, 287 (2003).

\bibitem[Singh et al.(2006)]{SNS06} Singh, A., A. Nir, and R. Semiat, ``Free-surface flow of concentrated suspensions," Int. J. Multiphase Flow {\bf 32}, 775 (2006).

\bibitem[Sierou and Brady(2002)]{SB02} A. Sierou and J.F. Brady, J. Rheol. {\bf 46} (5), 1031 (2002).

\bibitem[Smith et al.(2010)]{SBCB10} Smith, M.I., R. Besseling, M.E. Cates, and V. Bertola, ``Dilatancy in the flow and fracture of stretched colloidal suspensions,: Nature communications {\bf 1}, 114 (2010).

\bibitem[Sperl(2006)]{Sperl06} Sperl, M., ``Experiments on corn pressure in silo cells -- translation and comment of Janssen's paper from 1895," Granular Matter {\bf 8}, 59 (2006).

\bibitem[Tian et al.(2011)]{Ti11}Tian Y., M. Zhang, J. Jiang, N. Pesika, H. Zeng, J. Israelachvili, Y. Meng, and S. Wen,  ``Reversible shear thickening at low shear rates of electrorheological fluids under electric fields," Phys. Rev. E 83, 011401 (2011).


\bibitem[Trappe et al.(2001)]{TPCSW01} Trappe, V., V. Prasad, L. Cipelletti, P.N. Segre, and D.A. Weitz, ``Jamming phase diagram for attractive particles," Nature {\bf 411} 772 (2001).

\bibitem[van Alsten and Granick(1988)]{VG88} Van Alsten, J, and S. Granick, ``Molecular tribometry of ultrathin liquid films", Phys. Rev. Lett. {\bf 61} No. 22, 2570 (1988).

\bibitem[von Kann et al.(2011)]{KSLM11} S. von Kann, J. H. Snoeijer, D. Lohse, and D. van der Meer ``Non-monotonic settling of a sphere in a cornstarch suspension," Phys. Rev. {\bf E} 84, 060401(R) (2011).

\bibitem[Waitukaitis and Jaeger(2012)]{WJ12} Waitukaitis, S.R., and H.M. Jaeger, ``Impact-activated solidification of dense suspensions via dynamic jamming fronts," Nature {\bf 487}, 205-209 (2012).

\bibitem[Waitukaitis et al.(2013)]{WRVJ13} Waitukaitis,S. R.,  L. K. Roth, V. Vitelli, and H. M. Jaeger, ``Dynamic Jamming Fronts," Europhysics Letters, in press (May, 2013).

\bibitem[Wagner and Brady(2009)]{WB09} Wagner, N.J. and J.F. Brady, ``Shear thickening in colloidal dispersions," Phys. Today, Oct. 2009, 27.

\bibitem[Weitz et al.(1987)]{WSBK87} Weitz, D.A., J.P. Stokes, R.C. Ball, and A.P. Kushnick, ``Dynamic capillary pressure in porous media: origin of the viscous-fingering length scale," Phys. Rev. Lett. {\bf 59} (26), 2967 (1987).

\bibitem[Xu et al.(2011)]{XGSXJZ11}Xu, Y. L. , X. L. Gong, Y. Q. Sun, S. H. Xuan, W. Q. Jiang, and Z. Zhang, ``Evolution of the initial hole in vertically vibrated shear thickening fluids," Physical Review E {\bf 83}, 056311 (2011).

\bibitem[Xu et al.(2012)]{XRD12}Xu, X., S.A. Rice, and A.R. Dinner, ``Relation between ordering and shear thinning in colloidal suspensions," Proc. Nat. Acad. Sci. {\bf 110} (10) 3771-3776 (2012).

\bibitem[Zhang et al.(2008)]{ZMB08} Zhang, J.,  T. Majmudar, and R. Behringer, ``Force chains in a two-dimensional granular pure shear experiment," Chaos {\bf 18}, 041107 (2008).

\bibitem[Zheng et al.(2013)]{ZSX13}Zheng, W., Y. Shi, and N. Xu, ``Signatures of shear thinning-thickening transition in dense athermal shear flows," submitted.

\end{thebibliography}
\end{document}